\documentclass{article}

\usepackage{PRIMEarxiv}
\usepackage[title]{appendix}
\usepackage{xspace}
\usepackage{amsmath}
\usepackage{listings}
\usepackage{float}
\usepackage[pdftex]{graphicx,color}
\usepackage{algpseudocode}
\usepackage{algorithm}
\usepackage{refcount}
\usepackage{multicol}
\usepackage{tikz,pgfplots}
\usepackage{subcaption}
\usepackage{hyperref}
\appendixtitleon

\graphicspath{{figs/}}

\tikzset{algpxIndentLine/.style={{draw=white,dashed}}}
\author{
  Gaurav Bhardwaj \\
  Indian Institute of Technology \\
  Hyderabad.\\
  \texttt{cs19resch11003@iith.ac.in} \\
  \And
  Sathya Peri \\
  Indian Institute of Technology\\
  Hyderabad.\\
  \texttt{sathya\_p@cse.iith.ac.in}\\
  \And
  Pratik Shetty\\
   Indian Institute of Technology\\
  Hyderabad.\\
   \texttt{ai21mtech12005@iith.ac.in} 
}

\newcommand{\punt}[1]{}
\newcommand{\cmnt}[1]{}
\newcommand{\ignore}[1]{}

\newcommand{\secref}[1]{Section~\ref{sec:#1}}

\newcommand{\cgds} {concurrent graph data structure\xspace}

\newcommand{\op} {operation\xspace}
\newcommand{\ops} {operations\xspace}
\newcommand{\mth} {method\xspace}
\newcommand{\cc} {correctness-criterion\xspace}

\newcommand{\maxt}{{\tt Max\_Thread}\xspace}
\newcommand{\vntp}{{\tt VERTEX NOT PRESENT}\xspace}
\newcommand{\entp}{{\tt EDGE NOT PRESENT}\xspace}

\newcommand{\insrt}{{\tt Insert}\xspace}
\newcommand{\delte}{{\tt Delete}\xspace}
\newcommand{\ep}{{\tt EDGE PRESENT}\xspace}

\newcommand{\eadd}{{\tt EDGE ADDED}\xspace}
\newcommand{\er}{{\tt EDGE REMOVED}\xspace}
\newcommand{\ef}{{\tt EDGE FOUND}\xspace}
\newcommand{\vap}{{\tt VERTEX ALREADY PRESENT}\xspace}
\newcommand{\vp}{{\tt VERTEX PRESENT}\xspace}
\newcommand{\vadd}{{\tt VERTEX ADDED}\xspace}
\newcommand{\vr}{{\tt VERTEX REMOVED}\xspace}
\newcommand{\ICDCN}{{\tt Obst-Free}\xspace}
\newcommand{\reports}{{\tt WF}\xspace}
\newcommand{\pani}{{\tt PANIGRAHAM}\xspace}

\newcommand{\isMarked}{\textsc{isMarked}\xspace}
\newcommand{\isactive}{\textsc{IsActive}\xspace}
\newcommand{\deactivate}{\textsc{Deactivate}\xspace}
\newcommand{\MarkedRef}{\textsc{MarkedRef}\xspace}
\newcommand{\unMarkedRef}{\textsc{UnMarkedRf}\xspace}

\newcommand{\vnode}{\textsc{VNode}\xspace}
\newcommand{\enode}{\textsc{ENode}\xspace}
\newcommand{\senode}{\textsc{Snap\_Enode}\xspace}
\newcommand{\svnode}{\textsc{Snap\_Vnode}\xspace}

\newcommand{\addv}{\textsc{AddVertex}\xspace}
\newcommand{\remv}{\textsc{RemoveVertex}\xspace}
\newcommand{\conv}{\textsc{ContainsVertex}\xspace}
\newcommand{\adde}{\textsc{AddEdge}\xspace}
\newcommand{\cone}{\textsc{ContainsEdge}\xspace}
\newcommand{\reme}{\textsc{RemoveEdge}\xspace}
\newcommand{\locv}{\textsc{LocV}\xspace}
\newcommand{\loce}{\textsc{LocE}\xspace}
\newcommand{\convpl}{\textsc{ConVPlus}\xspace}
\newcommand{\locc}{\textsc{LocC}\xspace}
\newcommand{\concpl}{\textsc{ConCPlus}\xspace}
\newcommand{\reportV}{\textsc{ReportVertex}\xspace}
\newcommand{\reportE}{\textsc{ReportEdge}\xspace}
\newcommand{\reconst}{\textsc{ReconstructionUsingReports}\xspace}
\newcommand{\takes}{\textsc{TakeSnapshot}\xspace}
\newcommand{\acqs}{\textsc{AcquireSnapCollector}\xspace}

\newcommand{\itr}{\textsc{Iterator}\xspace}
\newcommand{\blkrep}{\textsc{BlockFurtherReports}\xspace}

\newcommand{\snap}{\textsc{Snap}\xspace}
\newcommand{\BC}{\textsc{BetweenCentrality}\xspace}
\newcommand{\bc}{\textsc{BC}\xspace}

\newcommand{\Diameter}{\textsc{Diameter}\xspace}
\newcommand{\dia}{\textsc{Dia}\xspace}

\newcommand{\vreport}{{\tt VReport}\xspace} 
\newcommand{\ereport}{{\tt EReport}\xspace} 
 
\newcommand{\scollect}{{\tt SnapCollector}\xspace}
\newcommand{\psc}{\textsc{\emph{PSC}}\xspace}
\newcommand{\headv}{\textsc{\emph{head\_Vnode}}\xspace}
\newcommand{\vreports}{\textsc{\emph{V\_reports}}\xspace}
\newcommand{\ereports}{\textsc{\emph{E\_reports}}\xspace}
\newcommand{\activef}{\textsc{\emph{active}}\xspace}
\newcommand{\reconstd}{\textsc{\emph{reconstruct\_done}}\xspace}

\newcommand{\collectvnode}{\textsc{CollectVnode}\xspace}
\newcommand{\collectenode}{\textsc{CollectEnode}\xspace}
\newcommand{\readcollectedsv}{\textsc{ReadCollectedVnodes}\xspace}
\newcommand{\readcollectedse}{\textsc{ReadCollectedEnodes}\xspace}
\newcommand{\readvreports}{\textsc{ReadVreports}\xspace}
\newcommand{\readereports}{\textsc{ReadEreports}\xspace}
\newcommand{\snapvnode}{\textsc{Snap\_Vnode}\xspace}
\newcommand{\snapenode}{\textsc{Snap\_Enode}\xspace}

\newcommand{\lble} {linearizable\xspace}

\newcommand{\lf}{lock-free\xspace}
\newcommand{\wf}{wait-free\xspace}
\newcommand{\of}{obstruction-free\xspace}

\newcommand{\gp}{graph-point\xspace}
\newcommand{\gs}{graph-set\xspace}

\begin{document}

\title{Non-blocking Dynamic Unbounded Graphs with Wait-Free Snapshot}
% \author{
% Gaurav Bhardwaj\inst{1} \and
% Sathya Peri\inst{1} \and
% Pratik Shetty\inst{1}
% }

% % \authorrunning{F. Author et al.}
% \institute{Indian Institute of Technology, Hyderabad, India.}

\maketitle
\begin{abstract}
Graphs are arguably one of the most fundamental data-structure used in many domains such as block-chain, networks etc. Theoretically and practically, improving Graph performance is one of the most studied and omnipresent research problems. In this paper, we have implemented a dynamic unbounded concurrent graph which can perform the add, delete or lookup operations on vertices and edges concurrently. All these methods are lock-free and linearizable. On top of this, we have also implemented the wait-free graph snapshot algorithm. To the best of knowledge this is first wait-free implementation of snapshot on concurrent graphs. We have used the snapshot of the algorithm to calculate the diameter and between centrality. We have compared our implementation with its counterparts and outperformed them by a good margin.  This illustrates the efficiency of our snapshot method which is a generic method and can be used to perform other useful graph analytics \op{s}. 
\end{abstract}

\section{Introduction}
\label{sec:intro}
Graph data structure have several real-life applications such as blockchains, maps, machine learning applications, biological networks, social networks, etc. A paired entity relation in a graph displays the relationship and structure between the objects. Social networks, for instance, use graphs to depict user relationships, which aids in making suggestions, spotting trends, and forecasting user behaviour. Over other data structures like linked lists, hash tables, trees, etc., graphs have a significant advantage in terms of application domains, making graph problem solving a major area of research.

%The Moore's law prediction that the number of transistors on silicon chips will stop increasing \cite{shalf2020future} implies the beginning of a purportedly new era of computing systems. In this setting, parallel processing is a requirement for efficient execution. As a result, new algorithms for data structures that can benefit from such systems must be developed. Data-parallel methods can process the enormous amount of data in a scalable manner. Hence parallelization is evidently an unavoidable proposition for data processing. A number of concurrent data structures, including stacks \cite{stack_spaa_2010}, queues \cite{Barnes:LFDS:SPAA:1993,Herlihy+:OFDQue:icdcs:2003,Kogan+:WFQue:ppopp:2011,Shavit+:LFQ:DC:2008}, linked lists \cite{chatterjee2016help,Harris:NBList:disc:2001,Heller+:LazyList:PPL:2007,Timnat:WFLis:opodis:2012,Valois:LFList:podc:1995,Zhang+:NBUnList:disc:2013}, hash table \cite{Liu+:LFHash:PODC:2014,Michael:LFHashList:spaa:2002}, etc., have been developed to take advantage of the capabilities of current multicore systems. 

With multi-core systems becoming ubiquitous from cell phones to high end computation servers, new algorithms for data structures that can benefit from such systems must be developed. Concurrent execution evidently is an natural proposition for harnessing these multi-core systems. A number of concurrent data structures, including stacks \cite{stack_spaa_2010}, queues \cite{Barnes:LFDS:SPAA:1993,Herlihy+:OFDQue:icdcs:2003,Kogan+:WFQue:ppopp:2011}, linked lists \cite{chatterjee2016help,Harris:NBList:disc:2001,Heller+:LazyList:PPL:2007,Timnat:WFLis:opodis:2012,Zhang+:NBUnList:disc:2013}, hash table \cite{Liu+:LFHash:PODC:2014,Michael:LFHashList:spaa:2002}, etc., have been developed to take advantage of the capabilities of current multicore systems. 

An important requirement for concurrent data-structures and algorithms is correctness. \emph{Linearizability} \cite{herlihy1990linearizability} is a commonly accepted \cc for concurrent systems which broadly states that a method is said to be \lble if it takes effect instantaneously at a point between its invocation and response during any concurrent execution. 

%Several concurrent applications use blocking synchronization primitives such as locks and barriers. Lock-based system can lead bottlenecks in performance and hamper progress through deadlocks etc. As a result researchers have explored non-blocking progress conditions while maintaining correctness. Wait-Freedom \cite{Herlihy+:OnNatProg:opodis:2011} is the maximum progress guarantee, which ensures that each process is completed in a finite number of steps. Wait freedom implementation results in expensive implementation costs and poor performance. Lock freedom ensures that at least one process will complete in a finite number of steps and is more practical. With progress assurance, consistency is also a crucial requirement for concurrent operations. The most prominent consistency framework is linearizability \cite{herlihy1990linearizability}, which ensures that the operation has an atomic effect between its invocation and return \cite{kogan2012methodology}.

Several concurrent applications use blocking synchronization primitives such as locks and barriers. Lock-based system can lead bottlenecks in performance and hamper progress through deadlocks etc. As a result researchers have explored non-blocking progress conditions while maintaining correctness. In a \emph{lock-free} data structure, threads do not acquire locks, and at least one non-faulty thread is guaranteed to finish its method in a finite number of steps \cite{Herlihy+:OnNatProg:opodis:2011}. Therefore, lock-free data structures are highly scalable and naturally fault-tolerant. 

%In lock-free systems, the system as a whole makes progress. But this does not prevent a thread from \stfdm . \emph{\wf} algorithms guarantees that every \mth invocation by a thread finishes in a finite number of steps independent of the speed or the failure of other processes \cite{Herlihy+:nature::opodis:2011}. The bounded \wf guarantees that every invocation finishes its execution in a finite and bounded number of steps. 

Although in \lf systems, the system as a whole makes progress, some threads can starve. On the other hand, \emph{\wf} \cite{Herlihy:WFS:TPLS:1991} algorithms guarantee that every \mth invocation by a thread finishes in a finite number of steps independent of the speed or the failure of other processes \cite{Herlihy+:OnNatProg:opodis:2011}. The bounded \wf guarantees that every invocation finishes its execution in a finite and bounded number of steps. 

As opposed to \lf, \wf progress conditions an \emph{\of} condition is one in which a \mth completes when no other thread is concurrently executing \cite{Herlihy+:OnNatProg:opodis:2011}. 

%This leads researchers to execute the maximum progress guarantee in the concurrent data structure. 

\vspace{1mm}
\noindent
\textbf{Related Work on Concurrent Graphs:} There has been a lot of interest on concurrent graph implementations of late. Most of these implementations support two kinds of \op{s}: (a) \emph{\gp} \mth{s}, which are adding/removing/ looking-up vertices/edges on the graph. These \op{s} can be considered as operating on one (or two) vertex point of interest. (b) \emph{\gs} \mth{(s)}, which involves taking a partial or complete snapshot of the graph. This \op has to consider and collect several vertices. We use the term \gs and snapshot interchangeably. 

%Kallimanis et al. \cite{Kallimanis+:WFGraph:opodis:2015} proposed a concurrent dynamic bounded graph with dynamic traversals that supports dynamic operations of adding/removing vertices/edges (\gp). Chatterjee et al. \cite{Chatterjee+:NbGraph:ICDCN-19} proposed an unbounded concurrent graph structure that supports adding/removing/lookup on vertices/edges. 

Kallimanis et al. \cite{Kallimanis+:WFGraph:opodis:2015} proposed a concurrent dynamic bounded graph that supports dynamic \gp operations and a graph traversal (\gs \mth) all of which are \wf. Chatterjee et al. \cite{Chatterjee+:NbGraph:ICDCN-19} proposed an unbounded concurrent graph structure that supports \lf \gp \mth{s}. Their implementation also supported path reachability queries on the dynamic graph which was implemented by taking a \lble (possibly partial) snapshot (again a \gs \mth) of the graph  to check for reachability. However, their snapshot \mth is only \of and hence may never complete. 

%internally uses a \lf linked list to represent the graph as a modified adjacency list. and checking for path reachability on the snapshot constructed

Later, Chatterjee et al. refined the above implementation by using hash tables and BSTs to represent the graph \cite{Chatterjee2021NonBlockingDU}. Then using various partial snapshots of the graph, they performed more sophisticated graph analytics operations such as \emph{SSSP} (Single Source Shortest Path), \emph{BC} (Between Centrality), \emph{BFS} (Breadth First Search). They showed that how partial snapshots (a \gs \mth) can be effectively used for performing various graph analytics. They also achieved significant performance over the existing work. The snapshot \mth technique used by them is however only \of. 

%They used \of snapshot to  The modified implementation can do more sophisticated graph analytics operations SSSP (Single Source Shortest Path), BC (Between Centrality), and so on. This improved the performance of the \gp \mth{s}. 

%There are several  graph snapshots are necessary to perform increasingly complex graph analytics operations on concurrent dynamic graphs. Chatterjee et al. \cite{Chatterjee+:NbGraph:ICDCN-19,Chatterjee2021NonBlockingDU} performed this process using obstruction-free graph snapshot techniques. They use a complex double collect technique to take a snapshot of the graphs. 

%Much recent research on the concurrent data structure has been undertaken in linearizable snapshots. Jayanti \cite{Jayanti2005AnOM} offered one wait-free technique for a bounded array snapshot. Petrank and Timnat \cite{Petrank2013LockFreeDI} later extended to perform wait-free snapshots on unbounded structures like linked-lists, skip-lists etc. Wei et al. \cite{wei+ConstantPPOPP2021} recently exploited the concept of versioning to perform a wait-free constant-time snapshot. 

\subsection{Our Contribution} %Paper Summary
\label{subsec:contrb}

It can be seen that constructing (partial) snapshots of a dynamic, concurrent graph efficiently is an important problem which can be use for various graph analytics \op{s} as shown by \cite{Chatterjee2021NonBlockingDU}. Among the various concurrent graph structures proposed in the literature, none of them support \wf snapshot construction for unbounded graphs which can be used for graph analytics. 

In this paper, we address this shortcoming. We develop a concurrent graph structure that supports \wf snapshot construction while the \gp \mth{s} are \lf. To illustrate the usefulness of the snapshot constructed, we use it to compute analytics \op{s} Betweeness Centrality (\bc) and Diameter (\dia). 

Our solution is an extension Chatterjee et al's \cite{Chatterjee+:NbGraph:ICDCN-19} concurrent framework for unbounded graphs. We extend their \gp \mth{s} for constructing a \wf snapshot of the graph which is based on the snapshot algorithm of Petrank and Timnat \cite{Petrank2013LockFreeDI} developed for iterators. 

\vspace{1mm}
\noindent
\textbf{Roadmap: } In \secref{prelims}, we explained the data structure and the operations supported by our graph implementation. We explained our notion for maintaining the concurrent graph in \secref{design} and the algorithm to construct a \wf snapshot for performing graph analytics. In \secref{results} we have shown the  experimental findings comparing our implementation with its state-of-the-art counterparts. Our approach outperforms its contemporaries by two orders of magnitude for different workloads.

\section{Preliminaries and ADT}
\label{sec:prelims}
%\subsection{Preliminaries}
We created a concurrent lock-free graph data structure that maintains the Vertices and Edges in an adjacency list formats. Our adjacency list structure is inspired by Chatterjee et al's \cite{Chatterjee+:NbGraph:ICDCN-19} implementation.  The adjacency lists are maintained as \lf linked lists. One list is kept to store all the vertices $V$ in the graph. Each vertex $v \in V$ maintains a \lf edge list to keep all the edges beginning with the vertex $v$. For more details please refer to Figure \ref{fig:conGraph} in Appendix.

%We've also added more complex wait-free analytics procedures including Snapshots, Between Centrality, All Pair Shortest Route, and Diametre. 

%$G = (V, E)$ represents a \textit{graph}, where $V$ is the set of vertices and $E$ is the set of edges. As we will only address the directed graph in this work, an edge $e \in E$ is an ordered pair of vertices $u,v$, where $u$ is the source and $v$ is the destination vertex. We have also implemented some of the graph analytic operations outlined below:

%\noindent
In addition to the \gp \mth{s} of \cite{Chatterjee+:NbGraph:ICDCN-19}, our implementation supports the following \gs \mth{s}:

\begin{enumerate}
    \item \textbf{Snapshot}: Given a graph $G$, returns a consistent state of the graph.
    % \item \textbf{All Pair Shortest Path}: Given a graph $G$, returns the shortest path with respect to the total number of edges traversed for all the pair of vertices $u,v \in V$.
    \item \textbf{Diameter}: Given a graph $G$, returns the shortest path with respect to the total number of edges traversed for two farthest nodes from all pair of vertices $u,v \in V$.
    \item \textbf{Betweenness Centrality}: Given a graph $G$, returns a vertex which lies  most frequently in the shortest path of all pair of vertices $u,v \in V$.
\end{enumerate}

\subsection{The Abstract Data Type (ADT)}
We define an $ADT$ $\mathcal{A}$ to be the collection of operations: $\mathcal{A}$ = \addv, \remv, \conv,  \adde, \reme, \cone, \snap, \BC, \Diameter.

\begin{enumerate}
    \item \addv(v): adds a vertex $v$ to $V$ ($V \leftarrow V \cup v$) if $v$ $\notin$ $V$ and returns $\vadd$. If $v \in V$ then returns \vap.
    \item \remv(v): removes a vertex $v$ from $V$ if $v \in V$ and returns $\vr$. If $v \notin V$ then returns \vntp.
    \item \conv(v): returns \vp if $v \in V$ otherwise returns \vntp.
    \item \adde(u,v): returns \vntp if $u \notin V \vee v \notin V$. If edge $e(u,v) \in E$, it returns \ep otherwise, it adds an edge $e(u,v)$ to $E$ $(E \leftarrow E \cup e(u.v)$) and returns \eadd.
    \item \reme(u,v): returns \vntp if $u \notin V \vee v \notin V$. If edge $e(u,v) \notin E$, it returns \entp; otherwise, it removes the edge $e(u,v)$ from 
    $E$ $(E \leftarrow E - e(u,v))$ and returns \er.
    \item \cone(u,v): returns \vntp if $u \notin V \vee v \notin V$. If edge $e(u,v) \notin E$, it returns $\entp$ otherwise, it returns \ep.
    \item \snap : returns the previously described consistent snapshot of the graph. 
    \item \BC : returns the Between Centrality of Graph $G$ as described above.
    % \item \APSP :  returns the \textsc{All Pair Shortest Path}(\APSP) of Graph $G$ as mentioned above.
    \item
    \Diameter: returns Diameter of graph $G$ as mentioned above.
\end{enumerate}
% \paragraph*{The Abstract Data Type (ADT).} We consider an ADT $\mathcal{A}$ as a set of operations: $\mathcal{A}$ = $\{\insertADT(K,V)$, ~$\remove(K)$, ~$\search(K)$,  ~$\rangeQuery(K1,K2)\}$
% \section{Literature Review}
% \label{sec:literature}
% \pagebreak 

\section{Design and Algorithm}
\label{sec:design}

We utilised the same graph structure of adjacency lists with lock-free linked lists as Chatterjee et al. \cite{Chatterjee+:NbGraph:ICDCN-19} employed.
We have separated the operations into two categories for clarity: a) \gp \op and b) \gs \op. Graph-point operations are comparable to those implemented by Chatterjee et al. \cite{Chatterjee+:NbGraph:ICDCN-19}, with modest adjustments to allow for more advanced wait-free graph analytics procedures.
Graph-set operation necessitates a consistent snapshot of the graph, which is inspired by Timnak and Shavit's \cite{Timnat:WFLis:opodis:2012} iterative wait-free snapshot approach. 

% \vspace{-12mm}
We will discuss the essential changes for the \gp \ops with the implementation of Chatterjee et al. \cite{Chatterjee+:NbGraph:ICDCN-19}. Timnak and Shavit's\cite{Timnat:WFLis:opodis:2012} snapshot implementation for the linked list used the forwarding principle to get the consistent snapshot. Our graph implementation used the same idea for our \gs \ops.
\begin{figure}[ht]
\begin{multicols}{2}
\scriptsize
\begin{lstlisting}
class VNode{
  int k
  VNode* vnxt
  ENode* ehead }

class ENode{
  int l
  VNode* ptv
  ENode* enext }

class VReport{
  VNode* vnode
  int action 
  VReport* nextReport}

class EReport{
  ENode* enode
  VNode* source
  int action  
  EReport* nextReport}

class Snap_Enode{
  Snap_Enode* enext
  Enode* enode }

class Snap_Vnode{ 
  Vnode* vnode
  Snap_VNode* vnext
  Snap_Enode* head_Enode
  edge_status  
  iter_edge_status }

class SnapCollector{
    active = false
    V_Report* V_reports[Max_Threads]
    EReport* E_reports[Max_Threads]
    reconstruct_done = false
    Snap_Vnode* head_Vnode }
\end{lstlisting}
\end{multicols}
\caption{Datastructure Components.}
\label{Fig:DS}
\end{figure}
% \pagebreak 

% class Graph {
%     VNode head_Vnode
%     Op AddVertex(k)
%     Op RemoveVertex(k)
%     Op ContainsVertex(k)
%     Op AddEdge(k,l)
%     Op ContainsEdge(k,l)
%     Op RemoveEdge(k,l)
%     Op locV(v)
%     Op locE(v,k)
%     Op ConVPlus (k,l)
%     Op locC(v,k)
%     Op ConCPlus (k,l) }

  % Op ReportVertex(VNode*, action, tid)
  %   Op ReportEdge(ENode*, action, tid)
  %   Op TakeSnapshot()
  %   Op AcquireSnapCollector()
  %   Op CollectSnapshot()
  %   Op ReconstructionUsingReports()
  %   Op Iterator() 
  %   Op Deactivate();
  %   Op IsActive();
  %   Op CollectVnode(Vnode);
  %   Op CollectEnode(Snap_Vnode , Enode)
  %   Op ReadCollectedVnodes();
  %   Op ReadCollectedEnodes(Snap_Vnode);
  %   Op ReadVreports();
  %   Op ReadEreports();
  %   Op BlockFurtherReports();
\subsection{Graph Point Operations}
%\vspace{-4mm}
As previously stated, Chatterjee et al. \cite{Chatterjee+:NbGraph:ICDCN-19} influenced our \gp \op, while Timnam and Shavit's \cite{Timnat:WFLis:opodis:2012} inspired \gs \op. We used the lock-free linked list \cite{Harris:NBList:disc:2001} structure for defining the graph's nodes and edges; the data structure is shown in Figure 1. Vertices are linked lists, and each vertex is connected to the edge linked list. Vnode is the node structure of a linked list vertex node; Enode is the node structure of a linked list edge node. Because we forward the value to the concurrent ongoing snapshot operation for each \gp operation, we must modify the \gp \op compared to the version of Chatterjee et al. \cite{Chatterjee+:NbGraph:ICDCN-19}. So we'll cover one of the \gp operations \addv in detail; the others are comparable and covered in Appendix \ref{app:point}. 

% \vspace{5mm}
\paragraph{\addv$(v)$:} If vertex $v$ does not exist in the graph, this method inserts it. So we look for $v$ in the vertex list and attempt to add it using the $CAS$ if it is not there. First, we find the location where the vertex $v$ must be added to the vertex at line \ref{addv:srch}. If the vertex is already in the list, the report vertex insert is forwarded at line \ref{addv:rep_ins1} and \vap is returned at line \ref{addv:return_vap}. Otherwise, it creates a new Vnode and executes a $CAS$ operation. If the $CAS$ operation succeeds at line \ref{addv:cas}, the vertex is inserted, and you return \vadd after forwarding the report vertex insert to the snapshot operation. Please see algorithm \ref{fig:addv} for more algorithmic details.

All other \gp \ops work the same way as \addv. If we discover a vertex or an edge in any of the operations, we must forward the report insert to the snapshot operation. Similarly, if we discover a marked vertex or edge that has not been physically removed, we must forward the report delete to the snapshot process before performing the physical deletion. These reports ensure that the operations are accurate so that the snapshot is consistent and the LP of all other operations is right and within bounds. In section \ref{sec:proof}, we will go over this in depth while analyzing the correctness of our method. Due to space constraints, we have covered all of the other \gp \ops in depth in Appendix \ref{app:point}.
\subsection{Graph Snapshot Operation}
As previously stated, our graph snapshot is inspired by Timnak and Shavit's \cite{Timnat:WFLis:opodis:2012} iterator snapshot algorithm. We used the forwarding principle, where we forward the value as reports to the snapshot operation if some concurrent snapshot operation occurs. The snapshot procedure initially gathers all the graph elements by traversing all its components. Meanwhile, all concurrent \gp \ops transfer the values of the component they act on to the snapshot method utilising the forwarding principle. After gathering all the data, items from the graph are added or removed based on the reports obtained during that period to generate a consistent picture.

We added some data structures to implement the snapshot, as shown in Figure \ref{Fig:DS}. \ereport and \vreport are the report structures for the edge and vertex, respectively, that must be passed to the concurrent snapshot method. These structures hold the action, i.e. insert or remove, and the information about the associated edge and vertex. These records are subsequently used to reconstruct the snapshot. These reports are maintained as a linked list, so each report keeps a reference to the next report in the variable next.

\textbf{\senode} and \textbf{\svnode} are used to store a copy of the graph by the snapshot method, as \enode, and \vnode is used to store the main graph. They have the same structure as \enode and \vnode, except that they retain the vertex and edge records from the initial graph. Variables such as $edge\_status$ and $iter\_edge\_status$ are used to efficiently reconstruct the graph using the snapshot technique. The use of these variables will become apparent later in the reconstruction method description. 
\begin{figure}[!ht]
    \scriptsize
    \begin{multicols}{2}
    
    \begin{algorithmic}[1]
    % \setcounterref{ALG@line}{alg:last-step}

    \Procedure{\reportV}{$victim$, $action$, $tid$}
    \State $SC \leftarrow$ dereference($PSC$)
    \If{($SC$.IsActive())} 
        \State $temp \leftarrow$ V\_reports[$tid$]
        \State $report \leftarrow$ new VReport($victim$, $action$, $temp$)
        \label{alg:repv:new_report}
        \State CAS(V\_Reports[$tid$], $temp$, $report$)
        \label{alg:repv:cas_report}
    \EndIf
    \EndProcedure
    \newline

    \Procedure{\reportE}{$victim$, $action$, $tid$, $source$}
    \State $SC \leftarrow$ dereference($PSC$)
    \If{($SC$.IsActive())} 
        \State $temp \leftarrow $ E\_reports[$tid$]
        \State $report \leftarrow$ new EReport($victim$, $source$, $action$, $temp$)
        \label{alg:repe:newrep}
        \State CAS(E\_Reports[$tid$], $temp$, $report$)
        \label{alg:repe:cas}
    \EndIf
    \EndProcedure
    \newline

    \Procedure{\reconst}{}
    \State $V$[] $\leftarrow SC$.\readcollectedsv()
    \label{alg:reconst:readv}
    \State v\_reports[] $\leftarrow$ $SC$.\readvreports()    
    \label{alg:reconst:readvrep}
    \State A VNode $N$ belongs to Snapshot iff: 
        \State\textbf{(}($N$ has a reference in $V$[]) $\lor$ ($N$ has \insrt Report in v\_reports[])) $\land$ ($N$ does not have a \delte Report in v\_reports[] \textbf{)}
        \label{alg:reconst:check1}
    \State $curr\_V \leftarrow$  $SC$.head\_Vnode
    \State e\_reports[] $\leftarrow SC$.\readereports()
    % \label{alg:reconst:readrep}
    \label{alg:reconst:readerep}
    \State // 1st iteration
    \While{($curr\_V$ != Null $\land$ $\neg SC$.\reconstd)}
        \If{CAS($curr\_V$.edge\_status, IDLE, ACTIVE)}
            \label{alg:reconst:idleact}
            \State $E$[] $\leftarrow$ $SC$.\readcollectedse($V$)
            \State A ENode $M$ belongs to Snapshot iff : 
            \State \textbf{(}($M$ has a reference in $E$[]) $\lor$ ($M$ has INSERT Report in e\_reports[])\textbf{)} $\land$ ($M$ has no DELETE Report in e\_reports[] ) $\land$ (The DESTINATION Vnode of $M$ belongs to snapshot)
            \State CAS($curr\_V$.edge\_status , ACTIVE , DONE)
        \EndIf
        \State $curr\_V \leftarrow$ $curr\_V \rightarrow$vnext
    \EndWhile
    \State // 2nd iteration
    \State $curr\_V \leftarrow$  $SC$.head\_Vnode
    \While{$curr\_V$ != Null $\land$ $\neg SC$.\reconstd:}
        \If{$curr\_V$.edge\_status $\leftarrow$ ACTIVE}
            \State$E$[] = $SC$.\readcollectedse($V$)
            \State A ENode $M$ belongs to Snapshot iff : 
            \State \textbf{(}($M$ has a reference in $E$[]) $\lor$ ($M$ has INSERT Report in e\_reports[])\textbf{)} $\land$ ($M$ has no DELETE Report in e\_reports[] ) $\land$ (The DESTINATION Vnode of $M$ belongs to snapshot)
            \State CAS($curr\_V$.edge\_status , ACTIVE , DONE)
            \label{alg:reconst:actdone2}

        \EndIf
        \State $curr\_V$ = $curr\_V \rightarrow$vnext
    \EndWhile
    \State $SC$.\reconstd $=$ true       \label{alg:last-step3}
    \EndProcedure
    \algstore{snap_report}
    \newline
    \end{algorithmic}
    \end{multicols}
\caption{Snapashot operations for report and reconstruction}
\end{figure}
The \textbf{\scollect} object is used to communicate amongst all the concurrent snapshot threads and maintain the graph snapshot. By holding a reference to the head of the Snap\_Vnode pointer variable, the Snap collector object preserves the graph information acquired by the snapshot operation. It also includes a list of vertices and edges reported by each thread. All concurrent snapshots collaborate such that all these threads imitate a single snapshot procedure. We used the boolean variables active and \reconstd for this collective operation of the concurrent snapshot. Active determines whether a concurrent snapshot is in progress, whereas \reconstd is used to determine whether the graph reconstruction is complete.

Now let us discuss some important variables and operations which are involved in snapshot collection.
Here SC is used to indicate SnapCollector Object.
\begin{itemize}
    \item \psc:  A global pointer that points to the current SC Object.
    \item \headv: This holds the pointer to the first vertex of the ordered vertex list of the collected graph.
    \item \activef: A boolean value which indicates if any threads are taking a snapshot using the current SC. This field is \textit{true} when an SC object is created and \textit{false} after any thread has completed the graph vertices and edges iteration.

    \item \vreports: \textit{Vreports} is an array of the same size as the total number of threads, with each element used to keep the vertex report provided by the specific thread in the form of a linked list. All forwarded reports are added to the head of the list in their specific slot for thread in the array. 

    \item \ereports: This is similar to \textit{Vreports}, except the threads add reports when there is an edge updation.

    \item \reconstd: A boolean value indicates that some thread has already completed the graph reconstruction from the reports.
% \vspace{0.5cm}
    \item \isactive() / \deactivate() : \isactive method returns the value of boolean field \emph{active} and \deactivate updates \emph{FALSE} to \emph{active}.
    \item \collectvnode(): In coordination with all snapshot threads, traverses all graph vertices and stores them in the SC object using the \emph{Snap\_Vnode} reference.
% \vspace{0.5cm}
    % \item \collectvnode(\vnode) : SC Object maintains a list of collected vertices \emph{Snap\_Vnode} which can be accessed through the \emph{head\_vnode} and tail pointer that points to the last Snap\_Vnode that was added. The implementation is similar to the lock-free queue of Michael and Scott. 
    
     % It first creates a new Snap\_Vnode Object SV and checks whether the tail points to the end of the list. If not, it updates the tail pointer to its next. Now it checks if the key of the new tail is greater than or equal to the input Vnode key. If yes, then it returns the Vnode pointed by the tail. If not, it inserts the SV at the end and then updates the tail pointer accordingly. CAS is used to help perform these updates. The first CAS could fail. This could be because of two cases either some thread has added a new SV, in which case we return the vnode pointed by it, or the tail pointer is marked to indicate no further vnodes should be added. If the first CAS succeeds, irrespective of the outcome of 2nd CAS, we return the input Vnode.

     % This implementation enforces ascending order amongst the keys of the collected vertices. Furthermore, it avoids duplication such that no two collected vertices will have duplicate keys. Moreover, since we are assuming a limited number of keys for the graph, the number of times the CAS fails during insertion is also limited. Therefore making the complete process wait-free.
     % \vspace{0.5cm}
     \item \collectenode(\snapvnode): This function, like \collectvnode, gathers all the outgoing edges of the relevant vertex by traversing them collaboratively. Using the head\_Enode pointer in \textit{Snap\_Vnode}, collected edges are kept in the list maintained with each vertex of \textit{Snap\_Vnode}. 
     % \vspace{0.5cm}
     % \vspace{0.5cm}
     \item \readcollectedsv(): returns the list of \snapvnode which were collected during iterator operation.
     % \vspace{0.5cm}
     \item \readcollectedse(\snapvnode): Returns the list of \snapenode which are collected for the given \snapvnode.
     % \vspace{0.5cm}
     \item \readvreports(): Returns the \vreports field if it is not null else, collects all the Vertex Reports from each thread, stores the sorted reports in \vreports, and returns it. These reports are sorted by increasing the order of the key of the \vnode field of the report.
     % \vspace{0.5cm}
     \item \readereports() : Implementation is similar to \readvreports but for edge reports.
     % \vspace{0.5cm}
     \item \reportV(\vnode, action, tid): If the current SC Object is Active, it creates a new Vertex Report Object at line \ref{alg:repv:new_report} and uses CAS  to update the head of the Linked List of Vertex Report for the given thread tid at line \ref{alg:repv:cas_report}. 
    % \vspace{0.5cm}
           \begin{figure}[H]
    \scriptsize
    % \caption{An algo cont..}\label{alg:four}
    \begin{multicols}{2}
    
    \begin{algorithmic}[1]
    \algrestore{snap_report}

    \Procedure{\takes}{}
        \State $SC \leftarrow$ \acqs()
        \State $SC$.\itr()        
        \State $SC$.\deactivate()
        \label{alg:ts:deact}
        \State $SC$.\blkrep()
        \State $SC$.\reconst()        
    \EndProcedure
    \newline
    
    \Procedure{\acqs}{}
    \State $SC \leftarrow$ dereference$(PSC)$
    \If{($SC$ is not Null $\land$ $SC$.IsActive())}  
        \State \Return $SC$ 
    \EndIf
    \State $newSC \leftarrow$ NewSnapCollector() 
    \State CAS($PSC$, $SC$, $newSC$) 
    \label{alg:aqsc:cas}
    \State $newSC \leftarrow$ dereference$(PSC)$ 
        \State \Return $newSC$  
    \EndProcedure
    \newline
    \Procedure{\blkrep}{}
    \For{($tid$ in range(\maxt))} 
        \State $temp \leftarrow$ \unMarkedRef(E\_Reports[$tid$])
        \State CAS(E\_Reports[$tid$], $temp$, \MarkedRef($temp$))
        \State $temp \leftarrow$  \unMarkedRef(V\_Reports[$tid$])
        \State CAS(V\_Reports[$tid$], $temp$, \MarkedRef($temp$))
    \EndFor
    
    \EndProcedure
    \newline

    \Procedure{\itr}{}
    % \State $curr\_V$ = graph's head vertex
    % \While{$curr\_V \neq$ Null $\land$ $SC$.\isactive()}
    %     \If{$curr\_V$ not marked}
    %         $curr\_V$ = \State $SC$.\collectvnode($curr\_V$)
    \State{$SC$.\collectvnode()}
    \label{alg:itr:cvnode}
    %     \EndIf
    %     \State $curr\_V$ = $curr\_V$.vnext
    % \EndWhile
        % \State $SC$.BlockFurtherVnodes()
        \State $curr\_V \leftarrow$ $SC$.head\_Vnode

    \State // 1st iteration
    \While{($curr\_V \neq$ Null $\land$ $SC$.\isactive())}
        \If{(CAS($curr\_V$.iter\_edge\_status , IDLE , ACTIVE))}
        \label{alg:itr:cas1}
            % \State $curr\_E = $  $curr\_V$.head\_Enode
            \State $SC$.\collectenode($curr\_V$)
            \label{alg:itr:collect_Enode}
            % \While{$curr\_E \neq $ Null}
            %     \If{$curr\_E$ not marked} 
            %         $curr\_E$ = \State $SC$.\collectenode(
            %         \State$curr\_V$, $curr\_E$)
            %     \EndIf
            %     \State $curr\_E =$ $curr\_E$.next
            % \EndWhile
            % \State $SC$.BlockFurtherEnodes($curr\_V$)
            \State CAS($curr\_V$.iter\_edge\_status , ACTIVE , DONE)
        \EndIf
        \State $curr\_V$ $\leftarrow$ $curr\_V \rightarrow$vnext
        \label{alg:itr:curr_next1}
    \EndWhile
    \State // 2nd Iteration
    \State $curr\_V$ $\leftarrow$ $SC$.head\_Vnode
    \While{($curr\_V \neq$ Null $\land$ $SC$.\isactive())}
        \If{CAS($curr\_V$.iter\_edge\_status $=$ ACTIVE)}
            % \State $curr\_E =$  $curr\_V$.head\_Enode
            % \While{$curr\_E \neq $ Null}
                % \If{$curr\_E$ not marked} 
            \State $SC$.\collectenode($curr\_V$)
        % \EndIf
            % \State $curr\_E =$ $curr\_E$.next
            % \EndWhile
        \State $SC$.BlockFurtherEnodes($curr\_V$)
        \State CAS($curr\_V$.iter\_edge\_status , ACTIVE , DONE)
        \EndIf
    \State $curr\_V \leftarrow curr\_V \rightarrow$ vnext
    \EndWhile
    \label{alg:end:snapshot}
    \EndProcedure
     \algstore{snap_op}
    \newline
        \end{algorithmic}
    \end{multicols}
    \caption{Snapshot Operation}
\end{figure}
    \item \reportE(\enode, action, tid): This method is similar to \reportV, except it creates an Edge Report Object at line \ref{alg:repe:newrep} and updates the head of Edge Report for the thread tid at line \ref{alg:repe:cas}.
    % \vspace{0.5cm}
    \item \acqs(): Returns the current SC Object from \textit{PSC} if it is Active. Otherwise, build an SC Object and use \textit{CAS} at line \ref{alg:aqsc:cas} to change the \textit{PSC} to point to the new SC object. It retrieves the most recent value saved in \textit{PSC}, considering that other threads may have updated concurrently.
% \vspace{0.5cm}
    \item \itr(): \takes method calls this method to start the snapshot process by iterating through the graph and gathering all the unmarked vertices and edges. The \collectvnode procedure at line \ref{alg:itr:cvnode} gathers all of the graph's vertices by traversing them in coordination with all of the other threads executing the same function. 
    
    %Please see the appendix for a more detailed description of \collectvnode.  
    
    We now have a set of vertices. Each of these collected vertices is linked to a list of edges in the graph, which must also be collected. Each vertex's distinct list of edges can be handled individually. To optimize, we distribute the vertices among all concurrent threads executing snapshots so that each thread processes a separate edge list. The \textit{iter\_edge\_status} variable is used to assign vertices to each thread. We need two iterations to finish collecting the graph's edges. In the initial iteration, each thread will operate independently on each edge list. In the second iteration, the faster threads will assist the slower threads in iterating the edge list that still needs to be finished. 
 
    Specifically, We iterate over the collected Vertices in the first iteration. The thread that successfully changes the \textit{iter\_edge\_status} of curr\_V from IDLE to ACTIVE at line \ref{alg:itr:cas1} using CAS begins iterating its curr\_V's edge list. curr\_V is the \svnode Object that is currently being accessed. If CAS fails, the process continues to the next collected Vertex. If \enode curr\_E is not marked while iterating over the edge list, we add it to the collected edge list using the collectEnode function at line \ref{alg:itr:collect_Enode}. When it reaches the end of the edge list, it invokes the BlockFurtherEnodes function, which, as the name implies, prevents further addition of Enodes but just for the current\_V. It then uses CAS to change the iter\_edge\_status of curr\_V from ACTIVE to DONE. 

    The 2nd iteration makes the process wait-free, as even if some thread fails, other active threads can complete the operation successfully, provided at least one thread is active. Following this action, all graph items are traversed and gathered to the \scollect object. After collecting all of the elements from the graph, the \takes method sets the active field to false, preventing any additional reports from being recorded.
% \vspace{0.5cm}
    \item \blkrep(): Once the SC Object is deactivated, i.e., the active field is set to false. We need to block further addition of reports by other concurrent threads performing any updation on the graph structure. Two concurrent snapshot-collecting threads may collect different reports before reconstruction without blocking the nodes. This scenario would lead to inconsistency while updating the collected vertices or edges during reconstruction. Hence, we must block each report list to avoid inconsistent snapshots amongst threads. We do that by iterating through the thread ids and marking the head of edge and vertex reports for each thread using CAS. 
% \vspace{0.5cm}
    \item \reconst(): This technique completes the snapshot by rebuilding the graph based on changes reported by other threads throughout the \itr process. In reconstruction, we add or remove vertices/edges from the gathered vertices/edges based on the report of other threads' actions.
    
   At line \ref{alg:reconst:readv}, we get all of the gathered vertices \textit{V}, followed by all of the vertex reports at line \ref{alg:reconst:readvrep}. A VNode \textit{N} will be included in the final list of collected vertices if there is a \textit{Snap\_Vnode} amongst \textit{V} that refers \textit{N} or an insert report for \textit{N}, but there must be no delete reports that reference \textit{N} at line \ref{alg:reconst:check1}. So, if there are any \textsc{Remove} reports for N, we must \textsc{Remove} it regardless of the amount of \textsc{Insert} reports if it is part of the gathered vertices. N's address will be different for every new Vnode generated with the same key. 

     Now that we have the full list of gathered vertices, we must collect all of the edges using reports. As we observe, much like the iterator approach, the edge list of these vertices may be handled individually.
     
    We acquire all edge reports at line \ref{alg:reconst:readerep} and begin with the head of the gathered vertices \textit{curr\_V}. In the first iteration, each thread will attempt to change the edge state of curr\_V from IDLE to ACTIVE at line \ref{alg:reconst:idleact}. The first thread that successfully updates will work on the \textit{curr\_V} edge list. Other threads will go to the following Vertex. The requirement to insert or remove an Edge is identical to the one for the Vertex given before, except for every Edge that will be introduced or is currently existing. We determine if their destination \textit{Vnode} has a reference in the gathered vertices. If not, they are eliminated Edge if they have already been gathered. 
    
    Once a thread completes the first iteration, it is possible that some threads processing an edge list would have become slow or inactive. If that happens, the reconstruction of the snapshot will be incomplete. So to help such threads complete the faster threads in 2nd iteration, search by iterating through collected vertices and checking if any of them have the edge\_status field still ACTIVE. If yes, it processes the edges similar to the first iteration and, once it completes, updates the \textit{edge\_status} of the corresponding Vertex to DONE using CAS at line \ref{alg:reconst:actdone2}.
% \vspace{0.5cm}
    \item \takes () : Using the methods provided above, this procedure is used to capture a graph snapshot. An iterator is used to collect all of the elements by traversing the graph at first, and then a consistent graph is built using the reports obtained. 
\end{itemize}

\section{Correctness}
\definecolor{clr1}{RGB}{00,36,92}
\definecolor{clr3}{RGB}{188,80,144}
\definecolor{clr2}{RGB}{174,205,194}
\pgfplotsset{every axis/.append style={
                    label style={font=\bfseries},
                    tick label style={font=\tiny},
                    ybar=1,
                    grid style={draw=white},
                    ymin = 10000,
                     xtick={1,2,3,4,5},
                     ymode = log,
                    xticklabels = {13,26,52,78,104},    
                    ytick={10000,100000,1000000,10000000},
                    bar width=2.5,
                    yminorticks = false,
                    }}

\begin{figure}[hbt!]
    \centering
    % \vspace{-3mm}
    \def \exp_loc{Experiments/Exp1}
\begin{subfigure}[b]{0.33\textwidth}
    
    \begin{tikzpicture}
    \begin{axis}[
    width=\linewidth,
    xtick={1,2,3,4,5},
    xticklabels = {13,26,52,78,104},
    ytick={1000,10000,100000,1000000,10000000},
    ymin = 1000,
    % x label style={at={(axis description cs:0.5,-0.1)},anchor=north},
    % y label style={at={(axis description cs:-0.1,.5)},rotate=90,anchor=south},
    % ytick={0.5,1.0,1.5,2.0,2.5,3,3.5,4},
    % legend pos=outer north east,
    % legend entries={\small{REPORT},\small{ICDCN},\small{OPODIS}},
    % legend pos=north west,
    % legend style={nodes={scale=0.3, transform shape}},
    % grid style=dashed,
]

    \addplot[
        color=clr1,
        fill=clr1,
        ]
        table[ignore chars={(,)},col sep=comma] {\exp_loc/up_report.csv};
        % coordinates {
        % (1000,0.670373)(1500,1.00508)(2000,1.341778)(2500,1.675643)(3000,2.011006)(3500,2.348578)(4000,2.680916)
        % }
    \addplot[
        color=clr2,
        fill=clr2
        % mark=Mercedes star flipped,
        ]
        table[ignore chars={(,)},col sep=comma] {\exp_loc/up_icdcn.csv};
        % coordinates {
        % (1000,0.268939)(1500,0.3985193)(2000,0.552735)(2500,0.679917)(3000,0.818706)(3500,0.963621)(4000,1.090492)
        % };
    \addplot[
        color=clr3,
        fill=clr3
        % mark=Mercedes star flipped,
        ]
        table[ignore chars={(,)},col sep=comma] {\exp_loc/up_opodis.csv};
    
        \end{axis}
        
    \end{tikzpicture}
     % \caption[]{Update Intensive}
\caption{}
\end{subfigure}
    \hspace{-3em}
    \def \exp_loc{Experiments/Exp2}
\begin{subfigure}[b]{0.33\textwidth}
    \begin{tikzpicture}
    \begin{axis}[
    width=\linewidth,
    xtick={1,2,3,4,5},
    xticklabels = {13,26,52,78,104},
    legend entries={\small{\reports},\small{\ICDCN},\small{\pani}},
        legend pos=north west,
        legend style={nodes={scale=0.6, transform shape}},
        legend style = { legend columns = 3,anchor=north , legend style={at={(0.5,1.4)}} }]
]

\addplot[
    color=clr1,
    fill=clr1,
    ]
    table[ignore chars={(,)},col sep=comma] {\exp_loc/up_report.csv};
    % coordinates {
    % (1000,0.670373)(1500,1.00508)(2000,1.341778)(2500,1.675643)(3000,2.011006)(3500,2.348578)(4000,2.680916)
    % }
\addplot[
    color=clr2,
    fill=clr2
    % mark=Mercedes star flipped,
    ]
    table[ignore chars={(,)},col sep=comma] {\exp_loc/up_icdcn.csv};
    % coordinates {
    % (1000,0.268939)(1500,0.3985193)(2000,0.552735)(2500,0.679917)(3000,0.818706)(3500,0.963621)(4000,1.090492)
    % };
\addplot[
    color=clr3,
    fill=clr3
    % mark=Mercedes star flipped,
    ]
    table[ignore chars={(,)},col sep=comma] {\exp_loc/up_opodis.csv};

    \end{axis}
    
    \end{tikzpicture}
     % \caption[]{Update Intensive}
\caption{}
\end{subfigure}
    \hspace{-3em}
    \def \exp_loc{Experiments/Exp3}
\begin{subfigure}[b]{0.33\textwidth}
    \begin{tikzpicture}
    \begin{axis}[
    width=\linewidth,
    xtick={1,2,3,4,5},
    xticklabels = {13,26,52,78,104},
    % x label style={at={(axis description cs:0.5,-0.1)},anchor=north},
    % y label style={at={(axis description cs:-0.1,.5)},rotate=90,anchor=south},
    % ytick={0.5,1.0,1.5,2.0,2.5,3,3.5,4},
    % legend pos=outer north east,
    % legend entries={\smallEPORT},\small{ICDCN},\small{OPODIS}},
    legend pos=north west,
    legend style={nodes={scale=0.7, transform shape}},
]

\addplot[
    color=clr1,
    fill=clr1,
    ]
    table[ignore chars={(,)},col sep=comma] {\exp_loc/up_report.csv};
    % coordinates {
    % (1000,0.670373)(1500,1.00508)(2000,1.341778)(2500,1.675643)(3000,2.011006)(3500,2.348578)(4000,2.680916)
    % }
\addplot[
    color=clr2,
    fill=clr2
    % mark=Mercedes star flipped,
    ]
    table[ignore chars={(,)},col sep=comma] {\exp_loc/up_icdcn.csv};
    % coordinates {
    % (1000,0.268939)(1500,0.3985193)(2000,0.552735)(2500,0.679917)(3000,0.818706)(3500,0.963621)(4000,1.090492)
    % };
\addplot[
    color=clr3,
    fill=clr3
    % mark=Mercedes star flipped,
    ]
    table[ignore chars={(,)},col sep=comma] {\exp_loc/up_opodis.csv};

    \end{axis}
    
    \end{tikzpicture}
     % \caption[]{Update Intensive}
\caption{}
\end{subfigure}
    \vspace{-1em}
    \par\bigskip % force a bit of vertical whitespace
    \def \exp_loc{Experiments/Exp1}
\begin{subfigure}[b]{0.33\textwidth}

    \begin{tikzpicture}
    \begin{axis}[
        % width=4.2cm,height=4.5cm,
        % bar shift=1.5,
        % ytick style={draw=small},
        % scaled y ticks=false,
        % xlabel style = {font=\bfseries\boldmath}
        % xlabel={Number of Threads},
        % ylabel={Average Time [msec]},
        % ylabel style={ font=\tiny },
        % ylabel near ticks,
        % xmin=0, xmax=70,
        % ymin=0, ymax=2.5,
        width=\linewidth,
        ymin = 1000,
        %  xtick={1,2,3,4,5},
        %  ymode = log,
        % xticklabels = {13,26,52,78,104},    
        ytick={1000,10000,100000,1000000},
        % bar width=2.5,
        % legend entries={\small{REPORT},\small{ICDCN},\small{OPODIS}},
        % legend pos=north west,
        % legend style={nodes={scale=0.5, transform shape}},
        % legend style = { legend columns = 3,anchor=north , legend style={at={(0.5,1.4)}} }]
        % grid style=dashed,
    ]
    
    \addplot[
    color=clr1,
    fill=clr1,
    ]
    table[ignore chars={(,)},col sep=comma] {\exp_loc/look_report.csv};
    % coordinates {
    % (1000,0.670373)(1500,1.00508)(2000,1.341778)(2500,1.675643)(3000,2.011006)(3500,2.348578)(4000,2.680916)
    % }
\addplot[
    color=clr2,
    fill=clr2
    % mark=Mercedes star flipped,
    ]
    table[ignore chars={(,)},col sep=comma] {\exp_loc/look_icdcn.csv};
    % coordinates {
    % (1000,0.268939)(1500,0.3985193)(2000,0.552735)(2500,0.679917)(3000,0.818706)(3500,0.963621)(4000,1.090492)
    % };
\addplot[
    color=clr3,
    fill=clr3
    % mark=Mercedes star flipped,
    ]
    table[ignore chars={(,)},col sep=comma] {\exp_loc/look_opodis.csv};
        \end{axis}
        \end{tikzpicture}
         % \caption{Lookup Intensive}
\caption{}
\end{subfigure}
    \hspace{-3em}
    \def \exp_loc{Experiments/Exp2}
\begin{subfigure}[b]{0.33\textwidth}
    \begin{tikzpicture}
    \begin{axis}[
        width=\linewidth,
         xtick={1,2,3,4,5},
            xticklabels = {13,26,52,78,104},
        % ytick={0.5,1.0,1.5,2.0,2.5,3,3.5,4},
        % legend pos=outer north east,
        % grid style=dashed,
    ]
    
    \addplot[
    color=clr1,
    fill=clr1,
    ]
    table[ignore chars={(,)},col sep=comma] {\exp_loc/look_report.csv};
    % coordinates {
    % (1000,0.670373)(1500,1.00508)(2000,1.341778)(2500,1.675643)(3000,2.011006)(3500,2.348578)(4000,2.680916)
    % }
\addplot[
    color=clr2,
    fill=clr2
    % mark=Mercedes star flipped,
    ]
    table[ignore chars={(,)},col sep=comma] {\exp_loc/look_icdcn.csv};
    % coordinates {
    % (1000,0.268939)(1500,0.3985193)(2000,0.552735)(2500,0.679917)(3000,0.818706)(3500,0.963621)(4000,1.090492)
    % };
\addplot[
    color=clr3,
    fill=clr3
    % mark=Mercedes star flipped,
    ]
    table[ignore chars={(,)},col sep=comma] {\exp_loc/look_opodis.csv};
        \end{axis}
        \end{tikzpicture}
         % \caption{Lookup Intensive}
\caption{}
\end{subfigure}
    \hspace{-3em}
    \def \exp_loc{Experiments/Exp3}
\begin{subfigure}[b]{0.33\textwidth}
    \begin{tikzpicture}
    \begin{axis}[
        width=\linewidth,
         xtick={1,2,3,4,5},
            xticklabels = {13,26,52,78,104},
        % ytick={0.5,1.0,1.5,2.0,2.5,3,3.5,4},
        % legend pos=outer north east,
        % grid style=dashed,
    ]
    
    \addplot[
    color=clr1,
    fill=clr1,
    ]
    table[ignore chars={(,)},col sep=comma] {\exp_loc/look_report.csv};
    % coordinates {
    % (1000,0.670373)(1500,1.00508)(2000,1.341778)(2500,1.675643)(3000,2.011006)(3500,2.348578)(4000,2.680916)
    % }
\addplot[
    color=clr2,
    fill=clr2
    % mark=Mercedes star flipped,
    ]
    table[ignore chars={(,)},col sep=comma] {\exp_loc/look_icdcn.csv};
    % coordinates {
    % (1000,0.268939)(1500,0.3985193)(2000,0.552735)(2500,0.679917)(3000,0.818706)(3500,0.963621)(4000,1.090492)
    % };
\addplot[
    color=clr3,
    fill=clr3
    % mark=Mercedes star flipped,
    ]
    table[ignore chars={(,)},col sep=comma] {\exp_loc/look_opodis.csv};
        \end{axis}
        \end{tikzpicture}
         % \caption{Lookup Intensive}
\caption{}
\end{subfigure}
      \caption{\scriptsize{Performance of our implementation compared to its counterparts. x-axis: Number of threads. y-axis: Average Time taken in microseconds.(a) Read Heavy workload with snapshot, (b) Read Heavy workload with Diameter, (c) Read Heavy workload with Betweenness Centrality, (d) Update Heavy with snapshot, (e) Update Heavy with Diameter, (f) Update Heavy with Betweenness Centrality.} }
      \label{fig:res1}
\end{figure}
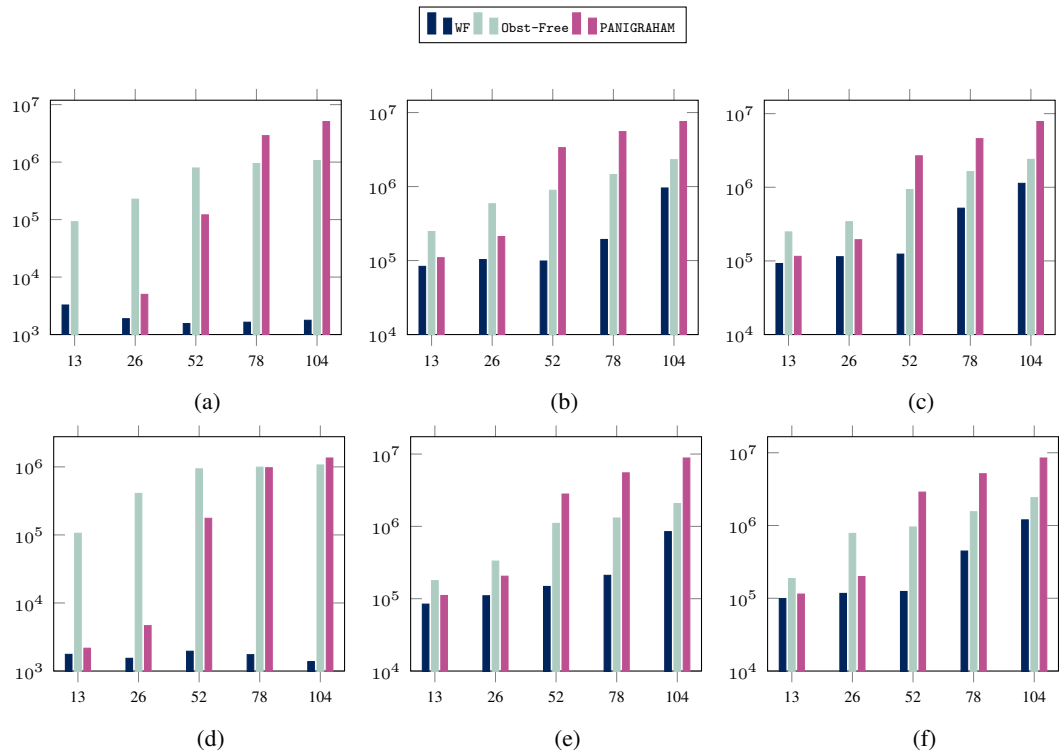

\label{sec:proof}
In this part, we will demonstrate the validity of our solution by demonstrating that all operations can be linearized. As previously stated, point operations are influenced by the implementation of Chatterjee et al. \cite{Chatterjee+:NbGraph:ICDCN-19}; hence LPs are the same as their prior implementation. So we will explain the LP's of the point operation in the Appendix \ref{app:corr} due to lack of space.

Similarly, the snapshot operation is linearizable, and before we explain the snapshot operation's linearization point, we'll review some points that will help you understand these LPs. If a point action is performed before the snapshot is deactivated, the point operation takes precedence over the snapshot. If P completes before S deactivates the snapshot at line \ref{alg:ts:deact}, then the snapshot operation will not return an older value for that vertex or edge.

Now we will demonstrate the linearization point for snapshot operations. The snapshot operation's linearization point occurs when we deactivate it at line \ref{alg:ts:deact}. All point operations completed up to that moment are gathered by the snapshot operation or documented through reports. We will now demonstrate how the LP of point operation varies in the presence of concurrent snapshot operation. Let S represent any snapshot operation, and P represent any concurrent point operation on any component i on the graph. 
\begin{enumerate}
    \item If S returns some older value than what P has updated or read, then the LP of P is just after the LP of S.
    \item If S returns newer or the value updated or written by P, then the LP of P doesn't change.
\end{enumerate}

\section{Experiments and Results}
\label{sec:results}

\pgfplotsset{every axis/.append style={
                    xticklabels = {2,4,6,8,10},    
                    }}

\begin{figure}[hbt!]
    \centering
    % \vspace{-3mm}
    \def \exp_loc{Experiments/Exp4}
   \begin{subfigure}[b]{0.33\textwidth}
    
    \begin{tikzpicture}
    \begin{axis}[
    width=\linewidth,
    xtick={1,2,3,4,5},
    xticklabels = {2,4,6,8,10},
     ymin = 1000,
        ytick={1000,10000,100000,1000000,10000000},
    % x label style={at={(axis description cs:0.5,-0.1)},anchor=north},
    % y label style={at={(axis description cs:-0.1,.5)},rotate=90,anchor=south},
    % ytick={0.5,1.0,1.5,2.0,2.5,3,3.5,4},
    % legend pos=outer north east,
    % legend entries={\small{REPORT},\small{ICDCN},\small{OPODIS}},
    legend pos=north west,
    legend style={nodes={scale=0.7, transform shape}},
]

\addplot[
    color=clr1,
    fill=clr1,
    ]
    table[ignore chars={(,)},col sep=comma] {\exp_loc/up_report.csv};
    % coordinates {
    % (1000,0.670373)(1500,1.00508)(2000,1.341778)(2500,1.675643)(3000,2.011006)(3500,2.348578)(4000,2.680916)
    % }
\addplot[
    color=clr2,
    fill=clr2
    % mark=Mercedes star flipped,
    ]
    table[ignore chars={(,)},col sep=comma] {\exp_loc/up_icdcn.csv};
    % coordinates {
    % (1000,0.268939)(1500,0.3985193)(2000,0.552735)(2500,0.679917)(3000,0.818706)(3500,0.963621)(4000,1.090492)
    % };
\addplot[
    color=clr3,
    fill=clr3
    % mark=Mercedes star flipped,
    ]
    table[ignore chars={(,)},col sep=comma] {\exp_loc/up_opodis.csv};

    \end{axis}
    
    \end{tikzpicture}
     % \caption[]{Update Intensive}
\caption{}
\end{subfigure}
    \hspace{-3.5em}
    \def \exp_loc{Experiments/Exp5}
\begin{subfigure}[b]{0.33\textwidth}
    \begin{tikzpicture}
    \begin{axis}[
    width=\linewidth,
    xtick={1,2,3,4,5},
    xticklabels = {2,4,6,8,10},
    legend entries={\small{\reports},\small{\ICDCN},\small{\pani}},
    legend pos=north west,
    legend style={nodes={scale=0.6, transform shape}},
    legend style = { legend columns = 3,anchor=north , legend style={at={(0.5,1.4)}} }]
]

\addplot[
    color=clr1,
    fill=clr1,
    ]
    table[ignore chars={(,)},col sep=comma] {\exp_loc/up_report.csv};
    % coordinates {
    % (1000,0.670373)(1500,1.00508)(2000,1.341778)(2500,1.675643)(3000,2.011006)(3500,2.348578)(4000,2.680916)
    % }
\addplot[
    color=clr2,
    fill=clr2
    % mark=Mercedes star flipped,
    ]
    table[ignore chars={(,)},col sep=comma] {\exp_loc/up_icdcn.csv};
    % coordinates {
    % (1000,0.268939)(1500,0.3985193)(2000,0.552735)(2500,0.679917)(3000,0.818706)(3500,0.963621)(4000,1.090492)
    % };
\addplot[
    color=clr3,
    fill=clr3
    % mark=Mercedes star flipped,
    ]
    table[ignore chars={(,)},col sep=comma] {\exp_loc/up_opodis.csv};

    \end{axis}
    
    \end{tikzpicture}
     % \caption[]{Update Intensive}
\caption{}
\end{subfigure}
    \hspace{-3.5em}
    \def \exp_loc{Experiments/Exp6}
\begin{subfigure}[b]{0.33\textwidth}
    
    \begin{tikzpicture}
    \begin{axis}[
    width=\linewidth,
    xtick={1,2,3,4,5},
    xticklabels={2,4,6,8,10},
    % x label style={at={(axis description cs:0.5,-0.1)},anchor=north},
    % y label style={at={(axis description cs:-0.1,.5)},rotate=90,anchor=south},
    % ytick={0.5,1.0,1.5,2.0,2.5,3,3.5,4},
    % legend pos=outer north east,
    % legend entries={\small{REPORT},\small{ICDCN},\small{OPODIS}},
    legend pos=north west,
    legend style={nodes={scale=0.7, transform shape}},
]

\addplot[
    color=clr1,
    fill=clr1,
    ]
    table[ignore chars={(,)},col sep=comma] {\exp_loc/up_report.csv};
    % coordinates {
    % (1000,0.670373)(1500,1.00508)(2000,1.341778)(2500,1.675643)(3000,2.011006)(3500,2.348578)(4000,2.680916)
    % }
\addplot[
    color=clr2,
    fill=clr2
    % mark=Mercedes star flipped,
    ]
    table[ignore chars={(,)},col sep=comma] {\exp_loc/up_icdcn.csv};
    % coordinates {
    % (1000,0.268939)(1500,0.3985193)(2000,0.552735)(2500,0.679917)(3000,0.818706)(3500,0.963621)(4000,1.090492)
    % };
\addplot[
    color=clr3,
    fill=clr3
    % mark=Mercedes star flipped,
    ]
    table[ignore chars={(,)},col sep=comma] {\exp_loc/up_opodis.csv};

    \end{axis}
    
    \end{tikzpicture}
     % \caption[]{Update Intensive}
\caption{}
\end{subfigure}
    \vspace{-1em}
    \par\bigskip % force a bit of vertical whitespace
    \def \exp_loc{Experiments/Exp4}
\begin{subfigure}[b]{0.33\textwidth}

    \begin{tikzpicture}
    \begin{axis}[
        width=\linewidth,
         xtick={1,2,3,4,5},
            xticklabels ={2,4,6,8,10},
            ymin = 1000,
        ytick={1000,10000,100000,1000000,10000000},
        % ytick={0.5,1.0,1.5,2.0,2.5,3,3.5,4},
        % legend pos=outer north east,
        % legend entries={\small{REPORT},\small{ICDCN},\small{OPODIS}},
        % legend pos=north west,
        % legend style={nodes={scale=0.4, transform shape}},
        % legend style = { legend columns = 3,anchor=north , legend style={at={(0.5,1.8)}} }]
        % grid style=dashed,
    ]
    
    \addplot[
    color=clr1,
    fill=clr1,
    ]
    table[ignore chars={(,)},col sep=comma] {\exp_loc/look_report.csv};
    % coordinates {
    % (1000,0.670373)(1500,1.00508)(2000,1.341778)(2500,1.675643)(3000,2.011006)(3500,2.348578)(4000,2.680916)
    % }
\addplot[
    color=clr2,
    fill=clr2
    % mark=Mercedes star flipped,
    ]
    table[ignore chars={(,)},col sep=comma] {\exp_loc/look_icdcn.csv};
    % coordinates {
    % (1000,0.268939)(1500,0.3985193)(2000,0.552735)(2500,0.679917)(3000,0.818706)(3500,0.963621)(4000,1.090492)
    % };
\addplot[
    color=clr3,
    fill=clr3
    % mark=Mercedes star flipped,
    ]
    table[ignore chars={(,)},col sep=comma] {\exp_loc/look_opodis.csv};
        \end{axis}
        \end{tikzpicture}
         % \caption{Lookup Intensive}
\caption{}
\end{subfigure}
    \hspace{-3.5em}
    \def \exp_loc{Experiments/Exp5}
\begin{subfigure}[b]{0.33\textwidth}
   \begin{tikzpicture}
    \begin{axis}[
        width=\linewidth,
         xtick={1,2,3,4,5},
            xticklabels ={2,4,6,8,10},
        % ytick={0.5,1.0,1.5,2.0,2.5,3,3.5,4},
        % legend pos=outer north east,
        % grid style=dashed,
    ]
    
    \addplot[
    color=clr1,
    fill=clr1,
    ]
    table[ignore chars={(,)},col sep=comma] {\exp_loc/look_report.csv};
    % coordinates {
    % (1000,0.670373)(1500,1.00508)(2000,1.341778)(2500,1.675643)(3000,2.011006)(3500,2.348578)(4000,2.680916)
    % }
\addplot[
    color=clr2,
    fill=clr2
    % mark=Mercedes star flipped,
    ]
    table[ignore chars={(,)},col sep=comma] {\exp_loc/look_icdcn.csv};
    % coordinates {
    % (1000,0.268939)(1500,0.3985193)(2000,0.552735)(2500,0.679917)(3000,0.818706)(3500,0.963621)(4000,1.090492)
    % };
\addplot[
    color=clr3,
    fill=clr3
    % mark=Mercedes star flipped,
    ]
    table[ignore chars={(,)},col sep=comma] {\exp_loc/look_opodis.csv};
        \end{axis}
        \end{tikzpicture}
         % \caption{Lookup Intensive}
\caption{}
\end{subfigure}
    \hspace{-3.5em}
    \def \exp_loc{Experiments/Exp6}
\begin{subfigure}[b]{0.33\textwidth}

    \begin{tikzpicture}
    \begin{axis}[
        width=\linewidth,
         xtick={1,2,3,4,5},
        xticklabels={2,4,6,8,10},
        % ytick={0.5,1.0,1.5,2.0,2.5,3,3.5,4},
        % legend pos=outer north east,
        % grid style=dashed,
    ]
    
    \addplot[
    color=clr1,
    fill=clr1,
    ]
    table[ignore chars={(,)},col sep=comma] {\exp_loc/look_report.csv};
    % coordinates {
    % (1000,0.670373)(1500,1.00508)(2000,1.341778)(2500,1.675643)(3000,2.011006)(3500,2.348578)(4000,2.680916)
    % }
\addplot[
    color=clr2,
    fill=clr2
    % mark=Mercedes star flipped,
    ]
    table[ignore chars={(,)},col sep=comma] {\exp_loc/look_icdcn.csv};
    % coordinates {
    % (1000,0.268939)(1500,0.3985193)(2000,0.552735)(2500,0.679917)(3000,0.818706)(3500,0.963621)(4000,1.090492)
    % };
\addplot[
    color=clr3,
    fill=clr3
    % mark=Mercedes star flipped,
    ]
    table[ignore chars={(,)},col sep=comma] {\exp_loc/look_opodis.csv};
        \end{axis}
        \end{tikzpicture}
         % \caption{Lookup Intensive}
\caption{}
\end{subfigure}
      \caption{\scriptsize{Performance of our implementation with varying percentage of graph analytics operation . x-axis: percentage of graph analytics operation. y-axis: Average Time taken in microseconds.(a) Read Heavy workload with snapshot, (b) Read Heavy workload with Diameter, (c) Read Heavy workload with Betweenness Centrality, (d) Update Heavy with snapshot, (e) Update Heavy with Diameter, (f) Update Heavy with Betweenness Centrality.} }
      \label{fig:res2}
\end{figure}
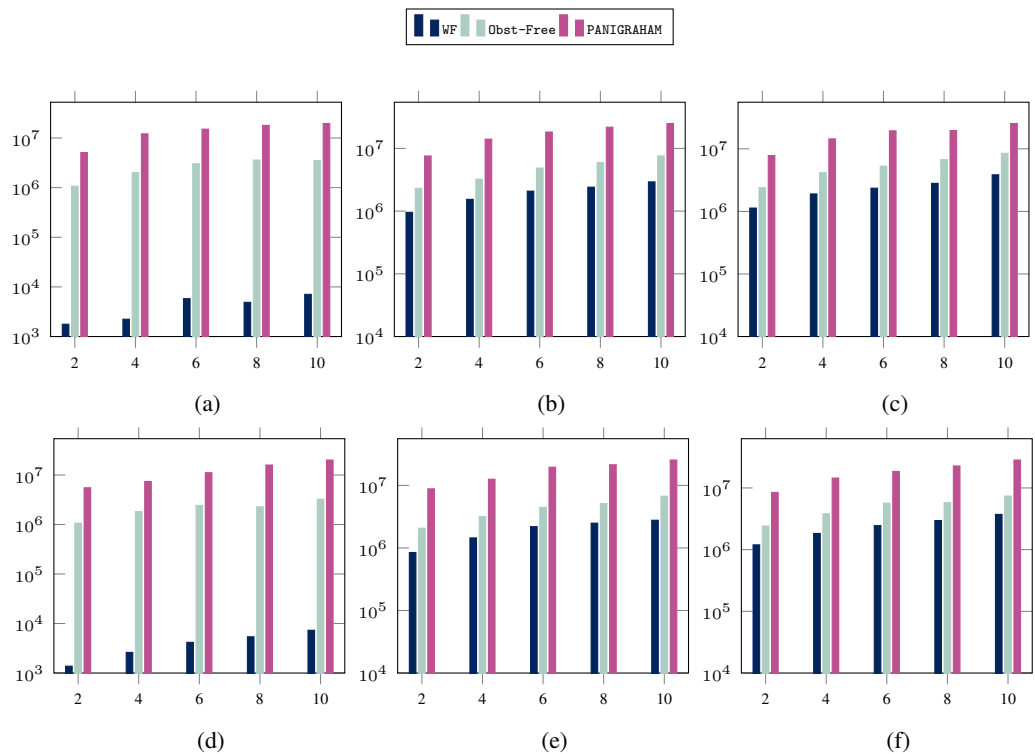
 
\paragraph{Platform Configuration:}
We conducted our experiments on a system with Intel(R) Xeon(R) Gold 6230R CPU packing 52 cores with a clock speed of 2.10GHz. There are 2 logical threads for each core, each with a private 32KB L1 and 1024KB L2 cache. The 36608KB L3 cache is shared across the cores. The system has 376GB of RAM and 1TB of  hard disk. It runs on a 64-bit Linux operating system. 

\paragraph{Experimental Setup:}All implementations\footnote{The code is available at \url{https://anonymous.4open.science/r/LF_Graph_WF_Iterator-0911/}.} are in C++ without garbage collection. We used Posix threads for multi-threaded implementation. We initially populated the graph with uniformly distributed synthetic data of 10K nodes and 20K edges for the experiments. In all our experiments, we have considered all the point operations \addv, \remv, \conv, \adde, \reme, \cone from ADT and one of the graph analytics operations from \snap, \BC and \Diameter. Each thread will execute for at least t seconds. During the execution, each thread randomly selects the next operation based on the workload distribution. Thus all the threads combined will perform n such operations within t secs. The evaluation metric used is Average Time, for which we consider the Average Time taken to complete each operation. We measure the Average Time w.r.t (i). Increasing spawned threads (ii) Increasing the workload to critical operations(Snapshot/Diameter/Betweenness Centrality) (iii) Multiple Snap graph Datasets.

\paragraph{Workload Distribution :}The distribution is over the following ordered set of Operations (AddVertex, RemoveVertex, ContainsVertex, AddEdge, RemoveEdge, ContainsEdge, and Critical Operation(Snapshot/Diameter/Betweenness Centrality). 
\begin{enumerate}
    \item Read Heavy Workload : 3\%, 2\%, 45\%, 3\%, 2\%, 45\% , 2\%
    \item Update Heavy Workload: 12\%, 13\%, 25\%, 13\%, 12\%, 25\% , 2\%
\end{enumerate}
% In one of the experiments, we measure the Average Time against the following Critical Section Workload : (2\%, 4\%, 6\%, 8\%, and 10\%), in which case we reduce the ContainsEdge and ContainsVertex workload by the same factor.

\paragraph{Algorithms :} We compare our wait-free Snapshot/Diameter/Betweenness Centrality approaches to the obstruction-free implementation of the same operations using Chatterjee et al. \cite{Chatterjee+:NbGraph:ICDCN-19} , and Chatterjee et al. \cite{Chatterjee2021NonBlockingDU}. We have named them \ICDCN and \pani, respectively, and our approach as \reports.
\pgfplotsset{every axis/.append style={
                    xticklabels = {p2p-Gnutella08,p2p-Gnutella04,p2p-Gnutella25,p2p-Gnutella24,p2p-Gnutella30},    
                    xticklabel style={rotate=30,anchor=east,}
                    }}
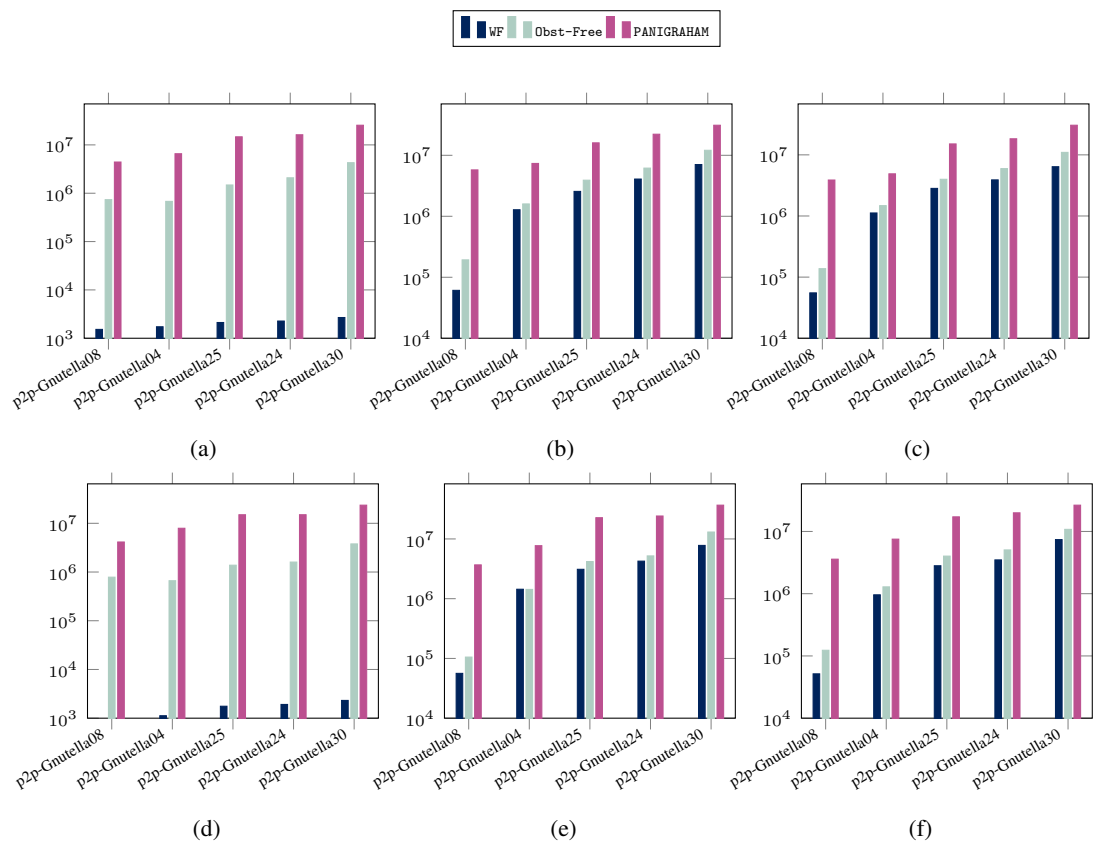
\begin{figure}[H]
    \centering
    % \vspace{-3mm}
    \def \exp_loc{Experiments/Exp7}
\begin{subfigure}[b]{0.33\textwidth}
    
    \begin{tikzpicture}
    \begin{axis}[
    width=\linewidth,
    xtick={1,2,3,4,5},
    ytick={1000,10000,100000,1000000,10000000},
    ymin = 1000,
    % x label style={at={(axis description cs:0.5,-0.1)},anchor=north},
    % y label style={at={(axis description cs:-0.1,.5)},rotate=90,anchor=south},
    % ytick={0.5,1.0,1.5,2.0,2.5,3,3.5,4},
    % legend pos=outer north east,
    % legend entries={\small{REPORT},\small{ICDCN},\small{OPODIS}},
    % legend pos=north west,
    % legend style={nodes={scale=0.3, transform shape}},
    % grid style=dashed,
]

    \addplot[
        color=clr1,
        fill=clr1,
        ]
        table[ignore chars={(,)},col sep=comma] {\exp_loc/up_report.csv};
        % coordinates {
        % (1000,0.670373)(1500,1.00508)(2000,1.341778)(2500,1.675643)(3000,2.011006)(3500,2.348578)(4000,2.680916)
        % }
    \addplot[
        color=clr2,
        fill=clr2
        % mark=Mercedes star flipped,
        ]
        table[ignore chars={(,)},col sep=comma] {\exp_loc/up_icdcn.csv};
        % coordinates {
        % (1000,0.268939)(1500,0.3985193)(2000,0.552735)(2500,0.679917)(3000,0.818706)(3500,0.963621)(4000,1.090492)
        % };
    \addplot[
        color=clr3,
        fill=clr3
        % mark=Mercedes star flipped,
        ]
        table[ignore chars={(,)},col sep=comma] {\exp_loc/up_opodis.csv};
    
        \end{axis}
        
    \end{tikzpicture}
     % \caption[]{Update Intensive}
\caption{}
\end{subfigure}
    \hspace{-3em}
    \def \exp_loc{Experiments/Exp8}
\begin{subfigure}[b]{0.33\textwidth}
    \begin{tikzpicture}
    \begin{axis}[
    width=\linewidth,
    xtick={1,2,3,4,5},
    legend entries={\small{\reports},\small{\ICDCN},\small{\pani}},
        legend pos=north west,
        legend style={nodes={scale=0.6, transform shape}},
        legend style = { legend columns = 3,anchor=north , legend style={at={(0.5,1.4)}} }]
]

\addplot[
    color=clr1,
    fill=clr1,
    ]
    table[ignore chars={(,)},col sep=comma] {\exp_loc/up_report.csv};
    % coordinates {
    % (1000,0.670373)(1500,1.00508)(2000,1.341778)(2500,1.675643)(3000,2.011006)(3500,2.348578)(4000,2.680916)
    % }
\addplot[
    color=clr2,
    fill=clr2
    % mark=Mercedes star flipped,
    ]
    table[ignore chars={(,)},col sep=comma] {\exp_loc/up_icdcn.csv};
    % coordinates {
    % (1000,0.268939)(1500,0.3985193)(2000,0.552735)(2500,0.679917)(3000,0.818706)(3500,0.963621)(4000,1.090492)
    % };
\addplot[
    color=clr3,
    fill=clr3
    % mark=Mercedes star flipped,
    ]
    table[ignore chars={(,)},col sep=comma] {\exp_loc/up_opodis.csv};

    \end{axis}
    
    \end{tikzpicture}
     % \caption[]{Update Intensive}
\caption{}
\end{subfigure}
    \hspace{-3em}
    \def \exp_loc{Experiments/Exp9}
\begin{subfigure}[b]{0.33\textwidth}
    \begin{tikzpicture}
    \begin{axis}[
    width=\linewidth,
    xtick={1,2,3,4,5},
    % xticklabels = {13,26,52,78,104},
    % x label style={at={(axis description cs:0.5,-0.1)},anchor=north},
    % y label style={at={(axis description cs:-0.1,.5)},rotate=90,anchor=south},
    % ytick={0.5,1.0,1.5,2.0,2.5,3,3.5,4},
    % legend pos=outer north east,
    % legend entries={\smallEPORT},\small{ICDCN},\small{OPODIS}},
    legend pos=north west,
    legend style={nodes={scale=0.7, transform shape}},
]

\addplot[
    color=clr1,
    fill=clr1,
    ]
    table[ignore chars={(,)},col sep=comma] {\exp_loc/up_report.csv};
    % coordinates {
    % (1000,0.670373)(1500,1.00508)(2000,1.341778)(2500,1.675643)(3000,2.011006)(3500,2.348578)(4000,2.680916)
    % }
\addplot[
    color=clr2,
    fill=clr2
    % mark=Mercedes star flipped,
    ]
    table[ignore chars={(,)},col sep=comma] {\exp_loc/up_icdcn.csv};
    % coordinates {
    % (1000,0.268939)(1500,0.3985193)(2000,0.552735)(2500,0.679917)(3000,0.818706)(3500,0.963621)(4000,1.090492)
    % };
\addplot[
    color=clr3,
    fill=clr3
    % mark=Mercedes star flipped,
    ]
    table[ignore chars={(,)},col sep=comma] {\exp_loc/up_opodis.csv};

    \end{axis}
    
    \end{tikzpicture}
     % \caption[]{Update Intensive}
\caption{}
\end{subfigure}
    \vspace{-1em}
    \par\bigskip % force a bit of vertical whitespace
    \def \exp_loc{Experiments/Exp7}
\begin{subfigure}[b]{0.33\textwidth}

    \begin{tikzpicture}
    \begin{axis}[
        % width=4.2cm,height=4.5cm,
        % bar shift=1.5,
        % ytick style={draw=small},
        % scaled y ticks=false,
        % xlabel style = {font=\bfseries\boldmath}
        % xlabel={Number of Threads},
        % ylabel={Average Time [msec]},
        % ylabel style={ font=\tiny },
        % ylabel near ticks,
        % xmin=0, xmax=70,
        % ymin=0, ymax=2.5,
        width=\linewidth,
        ymin = 1000,
        %  xtick={1,2,3,4,5},
        %  ymode = log,
        % xticklabels = {13,26,52,78,104},    
        ytick={1000,10000,100000,1000000,10000000},
        % bar width=2.5,
        % legend entries={\small{REPORT},\small{ICDCN},\small{OPODIS}},
        % legend pos=north west,
        % legend style={nodes={scale=0.5, transform shape}},
        % legend style = { legend columns = 3,anchor=north , legend style={at={(0.5,1.4)}} }]
        % grid style=dashed,
    ]
    
    \addplot[
    color=clr1,
    fill=clr1,
    ]
    table[ignore chars={(,)},col sep=comma] {\exp_loc/look_report.csv};
    % coordinates {
    % (1000,0.670373)(1500,1.00508)(2000,1.341778)(2500,1.675643)(3000,2.011006)(3500,2.348578)(4000,2.680916)
    % }
\addplot[
    color=clr2,
    fill=clr2
    % mark=Mercedes star flipped,
    ]
    table[ignore chars={(,)},col sep=comma] {\exp_loc/look_icdcn.csv};
    % coordinates {
    % (1000,0.268939)(1500,0.3985193)(2000,0.552735)(2500,0.679917)(3000,0.818706)(3500,0.963621)(4000,1.090492)
    % };
\addplot[
    color=clr3,
    fill=clr3
    % mark=Mercedes star flipped,
    ]
    table[ignore chars={(,)},col sep=comma] {\exp_loc/look_opodis.csv};
        \end{axis}
        \end{tikzpicture}
         % \caption{Lookup Intensive}
\caption{}
\end{subfigure}
    \hspace{-3em}
    \def \exp_loc{Experiments/Exp8}
\begin{subfigure}[b]{0.33\textwidth}
    \begin{tikzpicture}
    \begin{axis}[
        width=\linewidth,
         xtick={1,2,3,4,5},
            % xticklabels = {13,26,52,78,104},
        % ytick={0.5,1.0,1.5,2.0,2.5,3,3.5,4},
        % legend pos=outer north east,
        % grid style=dashed,
    ]
    
    \addplot[
    color=clr1,
    fill=clr1,
    ]
    table[ignore chars={(,)},col sep=comma] {\exp_loc/look_report.csv};
    % coordinates {
    % (1000,0.670373)(1500,1.00508)(2000,1.341778)(2500,1.675643)(3000,2.011006)(3500,2.348578)(4000,2.680916)
    % }
\addplot[
    color=clr2,
    fill=clr2
    % mark=Mercedes star flipped,
    ]
    table[ignore chars={(,)},col sep=comma] {\exp_loc/look_icdcn.csv};
    % coordinates {
    % (1000,0.268939)(1500,0.3985193)(2000,0.552735)(2500,0.679917)(3000,0.818706)(3500,0.963621)(4000,1.090492)
    % };
\addplot[
    color=clr3,
    fill=clr3
    % mark=Mercedes star flipped,
    ]
    table[ignore chars={(,)},col sep=comma] {\exp_loc/look_opodis.csv};
        \end{axis}
        \end{tikzpicture}
         % \caption{Lookup Intensive}
\caption{}
\end{subfigure}
    \hspace{-3em}
    \def \exp_loc{Experiments/Exp9}
\begin{subfigure}[b]{0.33\textwidth}
    \begin{tikzpicture}
    \begin{axis}[
        width=\linewidth,
         xtick={1,2,3,4,5},
            % xticklabels = {13,26,52,78,104},
        % ytick={0.5,1.0,1.5,2.0,2.5,3,3.5,4},
        % legend pos=outer north east,
        % grid style=dashed,
    ]
    
    \addplot[
    color=clr1,
    fill=clr1,
    ]
    table[ignore chars={(,)},col sep=comma] {\exp_loc/look_report.csv};
    % coordinates {
    % (1000,0.670373)(1500,1.00508)(2000,1.341778)(2500,1.675643)(3000,2.011006)(3500,2.348578)(4000,2.680916)
    % }
\addplot[
    color=clr2,
    fill=clr2
    % mark=Mercedes star flipped,
    ]
    table[ignore chars={(,)},col sep=comma] {\exp_loc/look_icdcn.csv};
    % coordinates {
    % (1000,0.268939)(1500,0.3985193)(2000,0.552735)(2500,0.679917)(3000,0.818706)(3500,0.963621)(4000,1.090492)
    % };
\addplot[
    color=clr3,
    fill=clr3
    % mark=Mercedes star flipped,
    ]
    table[ignore chars={(,)},col sep=comma] {\exp_loc/look_opodis.csv};
        \end{axis}
        \end{tikzpicture}
         % \caption{Lookup Intensive}
\caption{}
\end{subfigure}
      \caption{\scriptsize{Performance of our implementation with different Snap graph datasets. x-axis: Snap graph Datasets. y-axis: Average Time taken in microseconds. (a) Read Heavy workload with snapshot, (b) Read Heavy workload with Diameter, (c) Read Heavy workload with Betweenness Centrality, (d) Update Heavy with snapshot, (e) Update Heavy with Diameter, (f) Update Heavy with Betweenness Centrality.} }
      \label{fig:res3}
\end{figure}
% \vspace{-10mm}
\paragraph{Performance for various Graph Analytics Operation }In Figure \ref{fig:res1}, we compare the Average times of the algorithms under the two different workloads mentioned above. Initially, with Snapshot, and then we replace the Snapshot operation with Diameter and Betweenness Centrality(BC). In the case of Snapshot, our algorithm outperforms all its counterparts by up to two orders of magnitude. This occurs because, in our situation, if a new thread is required to execute Snapshot, it assists the current Snapshot if it is there and collaboratively finds the snapshot. Thus we see that the Average Time remains the same even with increasing active threads as more threads will be involved in creating a snapshot. On the other hand, in the obstruction-free algorithm, each thread creates its own independent Snapshot. Each thread performs the diameter and BC independently using the Snapshot in all three algorithms. Hence we see the Average time increasing with threads.

In Figure \ref{fig:res2}, we vary the percentage of the graph analytic operation from 2\% to 10\% for both Read Heavy and Update Heavy workloads. We set the number of threads to 104, and the proportion of graph analytics operations is calculated by equally reducing the percentage of \cone and \conv. We see that the performance of the \reports outperforms \ICDCN and \pani again with a significant margin. Because all threads collaborate, the average time of the snapshot process does not change significantly as the number of snapshot operations increases. \reports implementation exceeds their counterparts by a substantial amount in the case of BC and Diameter. However, the margin is smaller than in the snapshot, as all threads complete these activities independently after the snapshot. 

In Figure \ref{fig:res3}, we compare the performance of the algorithms against Snap datasets of the varying number of Nodes and Edges. We have used p2p-Fnutellla-
\noindent -08/04/25/24/30, ordered by the number of Nodes with max Nodes of around 36K and the workload of graph analytics operation to 2\%. This experiment shows that our algorithm's performance is not affected by the size of the graph.

% \label{sec:results}

\section{Conclusion and Future Work}
In this implementation, we have proposed the concurrent graph with a wait-free snapshot operation. Our algorithm performs all the point operations lock-free and the snapshot operations wait-free. It also outperforms all its counterparts by a significant margin. To our knowledge, our implementation provides the first wait-free snapshot operation on the graph. 

This implementation does not support the multithreaded model \cite{Bedin+:CAS-Burden:DISC:2021}, so it will remain a future work for us. Moreover, the current implementation performs point operation lock-free, so making it wait-free is another future work. 
\bibliography{Non_Blocking_Graph}

\begin{thebibliography}{10}

\bibitem{Barnes:LFDS:SPAA:1993}
Greg Barnes.
\newblock {A Method for Implementing Lock-free Shared-data Structures}.
\newblock In {\em SPAA}, pages 261--270, 1993.

\bibitem{Bedin+:CAS-Burden:DISC:2021}
Denis B\'{e}din, Fran\c{c}ois L\'{e}pine, Achour Most\'{e}faoui, Damien Perez,
  and Matthieu Perrin.
\newblock {Wait-Free CAS-Based Algorithms: The Burden of the Past}.
\newblock In Seth Gilbert, editor, {\em 35th International Symposium on
  Distributed Computing (DISC 2021)}, volume 209 of {\em Leibniz International
  Proceedings in Informatics (LIPIcs)}, pages 11:1--11:15, Dagstuhl, Germany,
  2021. Schloss Dagstuhl -- Leibniz-Zentrum f{\"u}r Informatik.
\newblock URL: \url{https://drops.dagstuhl.de/opus/volltexte/2021/14813}, \href
  {https://doi.org/10.4230/LIPIcs.DISC.2021.11}
  {\path{doi:10.4230/LIPIcs.DISC.2021.11}}.

\bibitem{Chatterjee2021NonBlockingDU}
Bapi Chatterjee, Sathya Peri, Muktikanta Sa, and Komma Manogna.
\newblock Non-blocking dynamic unbounded graphs with worst-case amortized
  bounds.
\newblock In {\em International Conference on Principles of Distributed
  Systems}, 2021.

\bibitem{Chatterjee+:NbGraph:ICDCN-19}
Bapi Chatterjee, Sathya Peri, Muktikanta Sa, and Nandini Singhal.
\newblock {A Simple and Practical Concurrent Non-blocking Unbounded Graph with
  Linearizable Reachability Queries}.
\newblock In {\em {ICDCN} 2019, Bangalore, India, January 04-07, 2019}, pages
  168--177, 2019.

\bibitem{chatterjee2016help}
Bapi Chatterjee, Ivan Walulya, and Philippas Tsigas.
\newblock {Help-optimal and Language-portable Lock-free Concurrent Data
  Structures}.
\newblock In {\em ICPP}, pages 360--369, 2016.

\bibitem{Harris:NBList:disc:2001}
Timothy~L. Harris.
\newblock {A Pragmatic Implementation of Non-blocking Linked-Lists}.
\newblock In {\em {DISC}}, pages 300--314, 2001.

\bibitem{Heller+:LazyList:PPL:2007}
Steve Heller, Maurice Herlihy, Victor Luchangco, Mark Moir, William N.~Scherer
  III, and Nir Shavit.
\newblock {A Lazy Concurrent List-Based Set Algorithm}.
\newblock {\em Parallel Processing Letters}, 17(4):411--424, 2007.

\bibitem{stack_spaa_2010}
Danny Hendler, Nir Shavit, and Lena Yerushalmi.
\newblock A scalable lock-free stack algorithm.
\newblock {\em J. Parallel Distrib. Comput.}, 70(1):1–12, jan 2010.

\bibitem{Herlihy:WFS:TPLS:1991}
Maurice Herlihy.
\newblock {Wait-free Synchronization}.
\newblock {\em ACM Trans. Program. Lang. Syst.}, 13(1):124--149, January 1991.

\bibitem{Herlihy+:OFDQue:icdcs:2003}
Maurice Herlihy, Victor Luchangco, and Mark Moir.
\newblock {Obstruction-Free Synchronization: Double-Ended Queues as an
  Example}.
\newblock In {\em {(ICDCS}}, pages 522--529, 2003.

\bibitem{Herlihy+:OnNatProg:opodis:2011}
Maurice Herlihy and Nir Shavit.
\newblock {On the Nature of Progress}.
\newblock In {\em {OPODIS}}, pages 313--328, 2011.

\bibitem{herlihy1990linearizability}
Maurice~P Herlihy and Jeannette~M Wing.
\newblock Linearizability: A correctness condition for concurrent objects.
\newblock {\em ACM Transactions on Programming Languages and Systems (TOPLAS)},
  12(3):463--492, 1990.

\bibitem{Kallimanis+:WFGraph:opodis:2015}
Nikolaos~D. Kallimanis and Eleni Kanellou.
\newblock {Wait-Free Concurrent Graph Objects with Dynamic Traversals}.
\newblock In {\em {OPODIS}}, pages 1--27, 2015.

\bibitem{Kogan+:WFQue:ppopp:2011}
Alex Kogan and Erez Petrank.
\newblock {Wait-Free Queues With Multiple Enqueuers and Dequeuers}.
\newblock In {\em {PPOPP}}, pages 223--234, 2011.

\bibitem{Liu+:LFHash:PODC:2014}
Yujie Liu, Kunlong Zhang, and Michael Spear.
\newblock {Dynamic-sized Nonblocking Hash Tables}.
\newblock In {\em PODC}, pages 242--251, 2014.

\bibitem{Michael:LFHashList:spaa:2002}
Maged~M. Michael.
\newblock {High Performance Dynamic Lock-Free Hash Tables and List-Based Sets}.
\newblock In {\em {SPAA}}, pages 73--82, 2002.

\bibitem{Petrank2013LockFreeDI}
Erez Petrank and Shahar Timnat.
\newblock Lock-free data-structure iterators.
\newblock In {\em International Symposium on Distributed Computing}, 2013.

\bibitem{Timnat:WFLis:opodis:2012}
Shahar Timnat, Anastasia Braginsky, Alex Kogan, and Erez Petrank.
\newblock {Wait-Free Linked-Lists}.
\newblock In {\em {OPODIS}}, pages 330--344, 2012.

\bibitem{Zhang+:NBUnList:disc:2013}
Kunlong Zhang, Yujiao Zhao, Yajun Yang, Yujie Liu, and Michael~F. Spear.
\newblock {Practical Non-blocking Unordered Lists}.
\newblock In {\em {DISC}}, pages 239--253, 2013.

\end{thebibliography}
\newpage
\appendix
% \noindent{\huge\bfseries APPENDIX\par}
\section{Adjacency List representation of Graph:}
\begin{figure}[ht]
    \centering
    \includegraphics[width=0.9\textwidth]{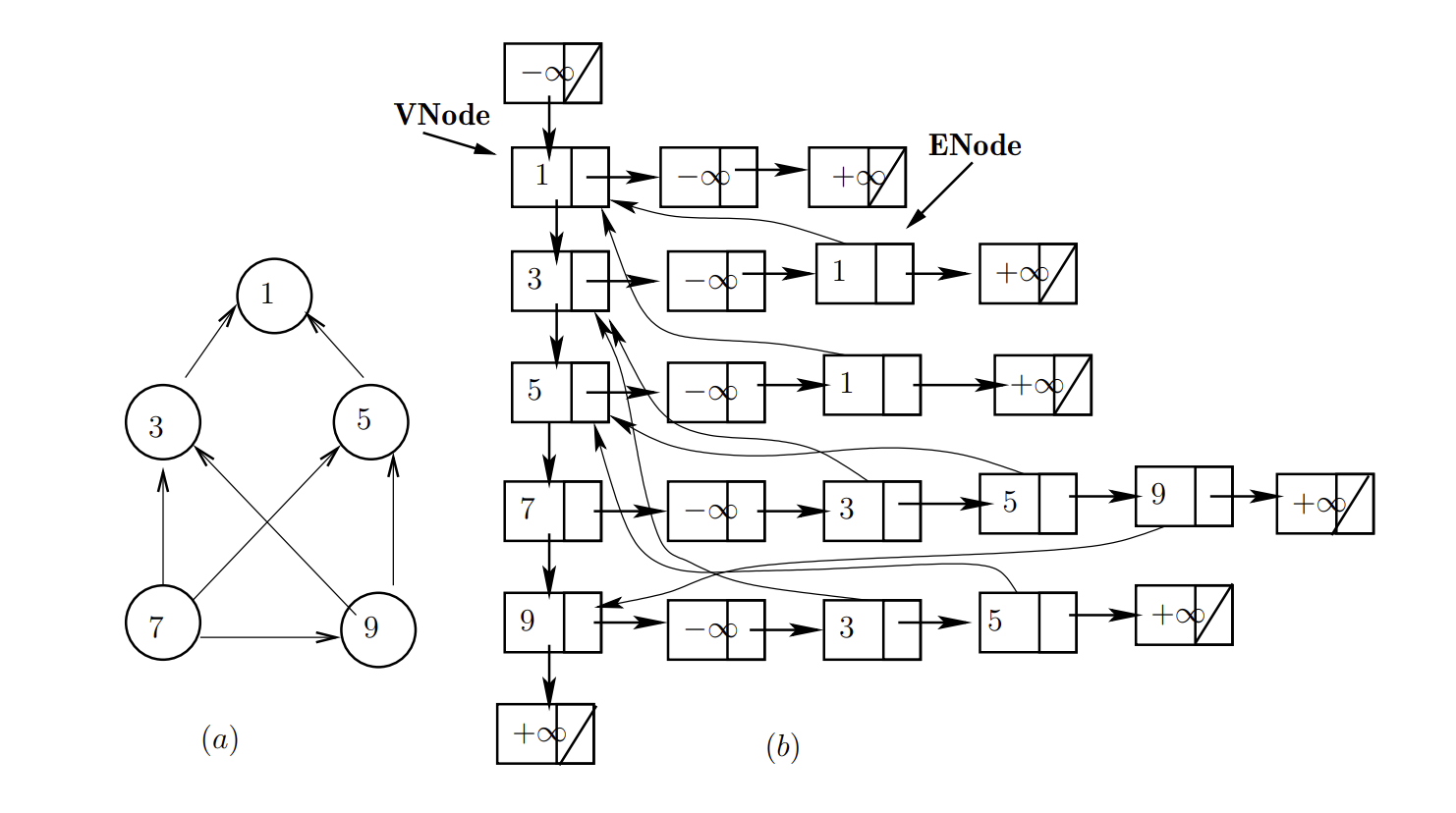}
    \caption{(a) A directed Graph (b) The \cgds for (a).}
    \label{fig:conGraph}
\end{figure}

This section will go through our graph adjacency list representation. The graph's adjacency list employs a lock-free linked list to facilitate concurrent \gp \op{s}. Like the \ICDCN implementation, our approach uses the \vnode to keep the lock-free list of vertices. Using the \enode pointer in \vnode, each vertex keeps a list of all the outgoing edges from that vertex. This \enode pointer in \vnode points to the beginning of the edge list. Similarly, a lock-free linked list is used to maintain the edge list. For more detail, please refer to Figure \ref{fig:conGraph}.

% \begin{figure}
% 	\captionsetup{font=footnotesize}
% 	%\captionsetup{font=tiny}
% 	\centerline{\scalebox{0.50}{\input{figs/conGraph.pdf_t}}}
%     %\vspace{-2mm}
% 	\caption{(a) A directed Graph (b) The \cgds for (a).}
% 	\label{fig:conGraph}
% \end{figure}

\section{Graph Point Operations}
\label{app:point}

\noindent In this section, we will demonstrate the point operation similar to \addv. The distinction between this operation and the \ICDCN, as indicated before in Section \ref{sec:design}, is that they obey the forwarding concept. When any operation discovers a marked vertex or edge, it transmits the delete report to the snapshot operation; similarly, if it successfully finds an unmarked vertex or edge, it forwards the insert report. We will now go over each procedure one by one.
\begin{figure}[ht]
    \scriptsize
    \begin{multicols}{2}
    \begin{algorithmic}[1]
    \algrestore{snap_op}
        \Procedure{\addv}{$k, tid$}
        \label{addv}
            \While{true}
                \State $\langle pv,cv \rangle \gets \locv(vh,k)$\
                \label{addv:srch}
                \If{($cv$.k = $k$)}
                \label{addv:if}
                    \State \reportV($cv$,\insrt, $tid$)
                    \label{addv:rep_ins1}
                    \State \Return \vap
                    \label{addv:return_vap}
                \Else
                    \State $nv \gets$   new \vnode($k$)
                    \label{addv:newnode}
                    \State $nv$.vnxt $\gets cv$
                    \label{addv:newnode_next}
                    \If{(CAS($pv$.vnxt, $cv, nv$))}
                    \label{addv:cas}
                        \State \reportV($nv$,\insrt, $tid$)
                        \label{addv:rep_ins2}
                        \State \Return \vadd
                        \label{addv:vadd}
                    \EndIf
                \EndIf
            \EndWhile
            \label{alg:last-step}
        \EndProcedure
        \algstore{addv}
    \end{algorithmic}
    \end{multicols}
\caption{Pseudo-code of Graph operation \addv}
\label{fig:addv}
\end{figure}
\begin{figure}
    \scriptsize
% \begin{algorithm}[H]
    % \caption{An algo}\label{alg:one}
    \begin{multicols}{2}

    \begin{algorithmic}[1]
        % \setcounterref{ALG@line}{alg:end:snapshot}
        \algrestore{addv}
        % \Procedure{\addv}{$k, tid$}
        % \label{addv}
        %     \While{true}
        %         \State $\langle pv,cv \rangle \gets \locv(vh,k)$\
        %         \label{addv:srch}
        %         \If{$cv$.k = $k$}
        %         \label{addv:if}
        %             \State \reportV($cv$,\insrt, $tid$)
        %             \label{addv:rep_ins1}
        %             \State \Return \vap
        %             \label{addv:return_vap}
        %         \Else
        %             \State $nv \gets$   new \vnode($k$)
        %             \label{addv:newnode}
        %             \State $nv$.vnxt $\gets cv$
        %             \label{addv:newnode_next}
        %             \If{CAS($pv$.vnxt, $cv, nv$)}
        %             \label{addv:cas}
        %                 \State \reportV($nv$,\insrt, $tid$)
        %                 \label{addv:rep_ins2}
        %                 \State \Return \vadd
        %                 \label{addv:vadd}
        %             \EndIf
        %         \EndIf
        %     \EndWhile
        %     \label{alg:last-step}
        % \EndProcedure
% \newline

        \Procedure{\remv}{$k, tid$}
            \While{true}
            \State $\langle     pv,cv \rangle \gets \locv(vh,k)$\
            \label{alg:remv:srch}
            \If{$cv.k \neq k$}
                \State \Return \vntp
                \label{alg:remv:vntp}
            \EndIf
            \State $cn \gets cv$.vnxt
            \If{$\neg \isMarked (cn)$}
                \If{CAS($cv$.vnxt, $cn$, $\MarkedRef(cn))$}
                \label{alg:remv:cas}
                    \State \reportV($cv$,\delte, $tid$)
                    \label{alg:remv:rep}
                    \If{CAS($pv$.vnxt, $cv, cn$)}
                    \label{alg:remv:cas1}
                        \State \textbf{break}
                    \EndIf
                \EndIf
            \EndIf
            \EndWhile
            \State \Return \vr
        \EndProcedure
\newline

        \Procedure{\conv}{$k, tid$}
        \State $cv \gets vh$.vnxt
        \While{$cv$.k $< k$} 
            \State $cv \gets \unMarkedRef(cv$.vnxt$)$
        \EndWhile
        \If{$cv .k = k \land \neg(\isMarked(cv)$}
        \State \reportV($cv$,\insrt, $tid$)
        \State \Return \vp
        \Else
            \If{$cv$.k = $k$ $\land$ \isMarked$(cv)$}
            \State \reportV($cv$,\delte, $tid$)
            \EndIf
            \State \Return \vntp
        \EndIf
        \EndProcedure
\newline
        
        \Procedure{\adde}{$k,l, tid$}
        \State $\langle u,v,st \rangle \gets \convpl (k,l)$
        \label{alg:adde:conv}
        \If{$st = false$}
                \State \Return \vntp
        \EndIf
        \While{true} 
            \If{$\isMarked(u)$}
                \State \reportV($u$, \delte, $tid$)
                \State \Return \vntp
            \EndIf
            \If{$\isMarked(v)$}
                \State \reportV($v$, \delte, $tid$)
                \State \Return \vntp
            \EndIf
            \State $\langle    pe,ce \rangle  \gets \loce(u$.enxt,$l)$\
            \If{$ce.l = l$}
                \State \reportE($ce$,\insrt, $tid$)
                \State \Return \ep 
            \EndIf
            \State $ne \gets$   new \enode($l$)
            \State $ne$.enxt $\gets ce$
            \State $ne$.ptv $\gets v$
            \If{CAS($pe$.enxt, $ce, ne$)}
            % \label{alg:adde:cas}
            \label{alg:adde:cas}
                    \State \reportE($ne$,\insrt, $tid$)
                    \State \Return \eadd
            \EndIf
        \EndWhile
        \State \textbf{end while}
        \EndProcedure
\newline

        \Procedure{\cone}{$k,l,tid$}
        \State $\langle    u,v,st \rangle  \gets \concpl(k,l)$\
        \If{$st = false$}
                \State \Return \vntp
            \EndIf
        \State $ce \gets u$.enxt
        \While{$ce.l < l$} 
            \State $ce \gets \unMarkedRef(ce$.enxt$)$
        \EndWhile
        \If{$ce.l = l$
        $\land$ $\neg$ 
        \isMarked ($u$)
        $\land$ $\neg$
        $\isMarked(v)$
        $\land$ $\neg$ \isMarked
        ($ce$)}
        \State \reportE($ce$,\insrt, $tid$)
        \State \Return \ef
        \Else
            \If{isMrkd($u$)}
            \State \reportV($u$,\delte, $tid$)
            \State \Return \vntp
            \EndIf

            \If{isMrkd($v$)}
            \State \reportV($v$,\delte, $tid$)
            \State \Return \vntp
            \EndIf
            \If{isMrkd($ce$)}
            \State \reportE($ce$,\delte, $tid$)
            \EndIf
            \State \Return \entp
        \EndIf
        \EndProcedure
\newline

        \Procedure{\reme}{$k,l,tid$}
        \State $\langle    u,v,st \rangle  \gets \convpl(k,l)$\
        \If{$st = false$}
                \State \Return \vntp
            \EndIf
        \While{true} 
            \If{\isMarked($u$)}
                \State \reportV($u$,\delte, $tid$)
                \State \Return \vntp
            \EndIf
            \If{\isMarked($v$)}
                \State \reportV($v$,\delte, $tid$)
                \State \Return \vntp
            \EndIf
            \State $\langle    pe,ce \rangle  \gets \loce(u$.enxt,$l)$\
            \If{$ce.l \neq l$}
                \State \Return \entp
            \EndIf
            \State $cnt \gets ce$.enxt
            \If{$\neg  \isMarked(cn)$}
                \If{CAS($ce$.enxt, $cnt, \MarkedRef(cnt$)}
                \label{alg:reme:cas}
                    \State \reportE($ce$,\delte, $tid$)
                    \If{CAS($pe$.enxt, $ce, cnt$)}
                        \State $break$
                    \EndIf
                \EndIf
            \EndIf
        \EndWhile
        \State \Return \er 
        \label{alg:end:graph_operations}
        \EndProcedure
        \algstore{graph_op}
\newline

    \end{algorithmic}
    \end{multicols}
% \end{algorithm}
\caption{Pseudo-code of Graph operations}
\end{figure}
% vnxt

\paragraph{\remv(v):} If the vertex $v$ exists in the graph, this procedure removes it. It uses \locv at line \ref{alg:remv:srch} to find the vertex's location $v$ in the vertex list, similar to \addv. If vertex does not exist, it returns \vntp at line \ref{alg:remv:vntp}; otherwise, at line \ref{alg:remv:cas}, it attempts to mark the vertex in the vertex list via $CAS$. If the $CAS$ operation is successful, the vertex is marked deleted, and a delete report is sent to the snapshot operation before the vertex is physically removed from the vertex list.

\paragraph{\conv(v):} This method searches the vertex list for the vertex v. If the vertex is present and unmarked, the insert vertex report is sent to the snapshot operation and returns \vp. Otherwise, if the vertex is marked, the remove vertex report is forwarded to the snapshot operation, and \vntp is returned. If the vertex is not found in the vertex list, it returns \vntp.

\paragraph{\adde(i,j):} If both vertices $i$ and $j$ exist and there is no edge from $i$ to $j$, this method inserts an edge from $i$ to $j$. At line \ref{alg:adde:conv}, it simply checks for the presence of the vertices $i$ and $j$ using \convpl; if either of them does not exist, it returns \vntp. If any vertices are marked, the remove vertex report is sent to the snapshot operation, and the \vntp is returned. The position of the edge($i,j$) in the edge list of vertex $i$ is then checked. We forward the insert edge report and return \ep if the edge is found in the edge list. Otherwise, we verify the references recorded before for the presence of vertices i and j once more. We send the delete vertex report to the snapshot operation if they are marked. If none of them is marked, it indicates that both vertices existed when we discovered the position for edge(i,j) at line \ref{alg:adde:conv}. If the edge already exists, we forward the edge insert report and return \ep; otherwise, the edge is added to the edge list through $CAS$ at line \ref{alg:adde:cas} and the edge insert report is forwarded to the snapshot operation and returns \eadd.
\begin{figure}
\scriptsize
    \begin{multicols}{2}
    
    \begin{algorithmic}[1]
    
        \algrestore{graph_op}
        \Procedure{\locv}{$k,l,tid$}
        \While{$true$}
        \label{alg:locv:while1}
        \State $pv \gets v$
        \State $cv \gets pv$.vnxt
            \While{$true$}
                \State $cn \gets cv$.vnxt
                \While{$\isMarked(cn)$ $\land$ $(cv$.k $< k)$}
                    \State \reportV($cv$,\delte, $tid$)
                    \If{$\neg$CAS($pv$.vnxt, $cv$, $\unMarkedRef$($cv$.vnxt)}
                        \State goto \ref{alg:locv:while1}
                    \EndIf
                    \State $cv \gets cn$
                    \State $cn \gets cv$.vnxt
                \EndWhile
                \If {$cv$.k $\geq k$}
                    \State \Return $(pv,cv)$
                \EndIf
            \State $pv \gets cv$
            \State $cv \gets cn$
        \EndWhile
        \EndWhile
        \EndProcedure
\newline

        \Procedure{\loce}{$v, k,tid$}
        \While{$true$}
        \label{alg:loce:while1}
            \State $pe \gets v$
            \State $ce \gets pe$.enxt
            \While{true}
                \State $cnt \gets ce$.enxt
                \State \vnode $vn \gets ce$.ptv
                \While{$\isMarked(vn)$ $\lor$ $\neg \isMarked (cnt)$}
                \label{alg:loce:while3}
                    \State \reportE($ce$,\delte, $tid$)
                    \If{$\neg$CAS($ce$.enxt, $cnt$, $\MarkedRef(cnt)$}
                        \State $goto$ $line$ \ref{alg:loce:while1}
                        \EndIf
                    \State \reportE($ce$,\delte, $tid$)
                    \If{$\neg$ CAS($pe.enxt$, $ce$, $cnt$)} 
                        \State$goto$ $line$ \ref{alg:loce:while1}
                        \EndIf
                    \State $ce \gets cnt$
                    \State $n \gets ce$.ptv
                    \State $cnt \gets \unMarkedRef(ce$.enxt$)$
                \EndWhile
                \While{$\isMarked(cnt)$}
                    \State \reportE($ce$,\delte, $tid$)
                    \If{$\neg$ CAS($pe$.enxt, $ce$, $cnt$)} 
                        \State $goto$ $line$ \ref{alg:loce:while1}
                    \EndIf
                    \State $ce \gets cnt$
                    \State $n \gets ce.ptv$
                    \State $cnt \gets$ $\unMarkedRef(ce.enxt)$
                \EndWhile
                \If{isMrkd($vn$)}
                    \State goto Line \ref{alg:loce:while3}
                \EndIf
                \If{$ce.l \geq k$}
                    \State \Return ($pe$,$ce$)
                \EndIf
                \State $pe \gets ce$
                \State $ce \gets cnt$
            \EndWhile
    \EndWhile
\EndProcedure
\newline

        \Procedure{\convpl}{$k, l$}
        \If{$k < l$}
            \State $\langle pv1, cv1 \rangle \gets \locv(vh, k)$
            \If{$cv1$.k $\neq k$}
                \State \Return $(Null,Null,false)$
            \EndIf
            \State $\langle pv2, cv2 \rangle \gets \locv(cv1, l)$
            \If{$cv2$.k $\neq l$}
                \State \Return $(Null,Null,false)$
            \EndIf
        \Else
            \State $\langle pv2, cv2 \rangle \gets \locv(vh, l)$
            \If{$cv2$.k $\neq l$}
                \State \Return $(Null,Null,false)$
            \EndIf
            \State $\langle pv1, cv1 \rangle \gets \locv(cv2, k)$
            \If{$cv1$.k $\neq k$}
                \State \Return $(Null,Null,false)$
            \EndIf
        \EndIf
    \State \Return $(cv1, cv2, true)$
    \EndProcedure
    \newline

        \Procedure{\locc}{$v, k$}
            \State $pv \gets v$
            \State $cv \gets p$.vnxt
            \While{$true$}
                \If{$cv$.k $\geq k$}
                    \State \Return $(p,c)$
                \EndIf
                \State $pv \gets cv$
                \State $cv \gets \unMarkedRef(cv$.vnxt$)$
            \EndWhile
    \EndProcedure
    \newline

    \Procedure{\concpl}{$k, l$}
        \If{$k < l$}
            \State $\langle pv1, cv1 \rangle \gets \locc(vh, k)$
            \If{$cv1$.k $\neq k$}
                \State \Return $(Null,Null,false)$
            \EndIf
            \State $\langle pv2, cv2 \rangle \gets \locc(cv1, l)$
            \If{$cv2$.k $\neq l$}
                \State \Return $(Null,Null,false)$
            \EndIf
        \Else
            \State $\langle pv2, cv2\rangle \gets \locc(vh, l)$
            \If{$cv2$.k $\neq l$}
                \State \Return $(Null,Null,false)$
            \EndIf
            \State $\langle pv1, cv1 \rangle \gets \locc(cv2, k)$
            \If{$cv1$.k $\neq k$}
                \State \Return $(Null,Null,false)$
            \EndIf
        \EndIf
    \State \Return $(cv1, cv2, true)$ \label{alg:last-step2}
    \EndProcedure
    \newline
    
        \end{algorithmic}
    \end{multicols}
    \caption{Graph Operations}
\end{figure}
\paragraph{\cone(i,j):} This method determines whether the edge($i,j$) exists in the graph. This method, like \adde, checks for the presence of both vertices; if neither exists, it simply returns \vntp. Otherwise, it looks for the edge in the vertex i's edge list; if the edge exists and neither of the vertices or edge is marked, we forward the insert edge report to the snapshot operation and return \ep. Otherwise, if any of the vertices or the edge is marked, we forward the delete report for that vertex or edge to the snapshot operation.

\paragraph{\reme(i,j):} This method removes the edge($i,j$) from the graph if it exists. Like \adde, it checks if the vertex or edge is absent and returns accordingly. If the edge exists, it simply marks it as deleted using $CAS$, forwards the delete edge report to the snapshot operation, and then physically deletes the edge from the edge list of vertex i.

\section{Correctness of Point operation}
\label{app:corr}
All the point operations specified in ADT in our graph implementations are linearizable. Now, depending on their return values, we will present the Linearization point of each of these procedures individually.

\begin{enumerate}
    \item \addv(k):
    \begin{enumerate}
        \item \vadd: Successful execution of \textit{CAS} operation at line \ref{addv:vadd}.
        \item \vap: Atomic read of vertex k in \locv.
    \end{enumerate}
    \item \remv(k):
    \begin{enumerate}
        \item \vr: Successful execution of $CAS$ at line \ref{alg:remv:cas}.
        \item \vntp: Atomic read of vertex p greater than k in the vertex list.
    \end{enumerate}
    \item \conv:
    \begin{enumerate}
        \item \vp: Atomic read of vertex k in the vertex list.
        \item \vntp: Atomic read of vertex p greater than k in the vertex list.
    \end{enumerate}
    \item \adde(i,j):
    \begin{enumerate}
        \item \vntp: Same as \conv.
        \item \ep: Atomic read of edge(i,j) in the edge list of vertex i.
        \item \eadd:
        \begin{enumerate}
            \item In case of no concurrent \remv the LP will be the successful execution of $CAS$ operation at line \ref{alg:adde:cas}.
            \item If \remv(i) or \remv(j) are running concurrently, the LP is either just before the LP of the first successful \remv operation or successful execution of line \ref{alg:adde:cas}, whichever occurs first. 
        \end{enumerate}
    \end{enumerate}
    \item \reme(i,j):
    \begin{enumerate}
        \item \vntp: Same as contains vertex.
        \item \entp: Atomic read of edge(i,k) in the edge list of vertex i where k is greater than j.
        \item \er: 
        \begin{enumerate}
            \item In case of no concurrent \remv the LP will be the successful execution of $CAS$ operation at line \ref{alg:reme:cas}.
            \item In the same way that \adde if \remv(i) or \remv(j) are both operating at the same time, the LP is either just before the LP of the first successful \remv operation or the successful execution of line \ref{alg:reme:cas}, whichever comes first. 
        \end{enumerate}
    \end{enumerate}
    \item \cone(i,j):
    \begin{enumerate}
        \item \ep: 
        \begin{enumerate}
            \item In case of no concurrent \remv the LP will be, the atomic read of edge(i,j) in the edge list of vertex i. 
            \item In case of concurrent \remv(i) or \remv(j) LP will be either just before the LP of first \remv operation or the atomic read of edge(i,j) in the edge list of vertex i,, whichever comes first.
        \end{enumerate}
        \item \entp: 
        \begin{enumerate}
            \item In case of no concurrent \remv the LP is atomic read of edge(i,k) in the edge list of vertex i where k is greater than j.
            \item LP will be either just before the LP of first \remv operation or atomic read of edge(i,k) in the edge list of vertex i where k is greater than j, whichever comes first.
        \end{enumerate}
        \item \vntp: same as \vntp for \conv.
    \end{enumerate}
\end{enumerate}
\end{document}